\documentclass[useAMS,usenatbib,usegraphicx]{mn2e}

\title[Bondi accretion in BCGs]{AGN jet power and feedback characterised
by Bondi accretion in brightest cluster galaxies}
\author[Y. Fujita et al.]{Yutaka
Fujita$^{1}$\thanks{E-mail: fujita@vega.ess.sci.osaka-u.ac.jp}, 
Nozomu Kawakatu$^{2}$, and
Isaac Shlosman$^{3,1}$
\\ 
$^{1}$Department of Earth and
Space Science, Graduate School of Science, Osaka University, 1-1
Machikaneyama-cho, \\ Toyonaka, Osaka 560-0043, Japan\\ 
$^{2}$Faculty of Natural Sciences, National Institute of Technology,
Kure College, 2-2-11 Agaminami, Kure, Hiroshima, 737-8506, Japan\\
$^{3}$Department of Physics and Astronomy, University of Kentucky,
Lexington, KY 40506-0055, USA}

\begin{document}

\date{Accepted 0000 December 00. Received 0000 December 00; in original form 0000 October 00}

\pagerange{000--000} \pubyear{0000}

\maketitle

\label{firstpage}

\begin{abstract}
 We propose a new method to estimate the Bondi (hot gas) accretion
 rates, $\dot M_{\rm B}$, onto the supermassive black holes (SMBHs) at
 the centres of elliptical galaxies. It can be applied even if the Bondi
 radius is not well-resolved in X-ray observations and it is difficult
 to measure the gas density and temperature there. This method is based
 on two simple assumptions: (1) hot gas outside the Bondi radius is in
 nearly a hydrostatic equilibrium, and (2)
 the gas temperature near the galaxy centre is close to the virial
 temperature. We apply this method to 28 bright elliptical
 galaxies in nearby galaxy clusters (27 of them are the brightest
 cluster galaxies; BCGs). We find a correlation between the Bondi
 accretion rates and the power of jets associated with the SMBHs over
 four orders of magnitude in $\dot M_{\rm B}$. For most galaxies, the
 accretion rates are large enough to account for the jet powers, which
 is in contrast with previous studies. Our results indicate that
 the feedback from the active galactic nuclei (AGN) correlates with
 the properties of the hot gas surrounding the SMBHs. We also find
 that more massive SMBHs in BCGs tend to have larger specific growth
 rates. This may explain the hyper masses ($\sim 10^{10}\: M_\odot$) of
 some of the SMBHs. Comparison between the accretion rates and the X-ray
 luminosities of the AGN suggests that the AGN in the BCGs are extremely
 radiatively inefficient compared with X-ray binaries in the Milky Way,
 even when their Eddington accretion ratio, $\dot M_{\rm B}/\dot M_{\rm
 Edd}$, exceeds 0.01. The corollary is that this ratio is not the only
 parameter which controls the radiative efficiency of the accretion
 flow.  Lastly, we find a tight correlation between the Bondi accretion
 rates and the X-ray luminosities of cool cores. Their relation is
 linear and the power generated by the Bondi accretion is large enough
 to compensate the radiative cooling of the cool cores. Although
 the `classical' Bondi accretion model is a greatly oversimplified one,
 the correlations  we find here demonstrate that the accretion onto the 
 SMBHs reflects broadly the properties of the Bondi accretion in some 
 time-averaged sense.

\end{abstract}

\begin{keywords}
accretion, accretion discs --- black hole physics --- galaxies: active
--- galaxies: jets --- X-rays: galaxies.
\end{keywords}

\section{Introduction}
\label{sec:intro}

Energetic feedback from active galactic nuclei (AGN) is thought to
prevent the cooling of hot gas in galaxies and clusters of galaxies,
delaying or suppressing star formation in these objects
\citep{bow06a,bes06a}. However, the mechanism that controls the level of
activity is not well known.  The AGN located at the centres of the
brightest cluster galaxies (BCGs) can play an important role in
preventing the development of ``cooling flows'', which would have
developed in the absence of the AGN feedback \citep{mcn07a}. However,
AGN feedback in cool cores of clusters is often subject to a global
thermal instability \citep[e.g.][]{fuj05b,mat06a}. Even if the thermal
instability does not develop with the aid of thermal conduction or
cosmic rays \citep{rus02a,guo08a,fuj13c}, the AGN must respond quickly,
on a dynamical or a sound crossing timescale in the core, to the
change of the environment (e.g. gas density and temperature) to maintain
the balance between the heating and the radiative cooling.

The Bondi accretion \citep{bon52a} is a promising route for the gas
supply to the central supermassive black holes (SMBHs) in the
BCGs. Since accretion rate depends on the density and temperature of the
surrounding hot gas (see equation~\ref{eq:dotMB}), the AGN activity can
respond accordingly to the change of the state of this gas. The AGN
feedback to the Bondi accretion has been detected by studying 9 nearby,
X-ray luminous elliptical galaxies (not all of them being BCGs) with
{\it Chandra X-Ray Observatory} \citep{all06a}; a correlation between
the Bondi accretion rates (inferred from the observed gas temperature
and density profiles, and SMBH masses) and the AGN power (in the form of
relativistic jets) has been shown to exist. Successive studies have
confirmed this correlation \citep{bal08a,vat10a}, although observational
uncertainties exist \citep{rus13b}.

For BCGs in general, the correlation has not been firmly established,
partly because observational uncertainties appear to be substantial due
to the large average distance to the BCGs. Moreover, apparent jet powers
seem to be insufficient to compensate for the radiative cooling of cool
cores of some clusters \citep{bir04a,raf06a}. As an alternative to the
Bondi hot gas accretion, cold gas accretion may work. In fact, cold gas
has been detected in many elliptical galaxies including BCGs
\citep[e.g.][]{edg01a,wer14a}, and some of them even have cold gas discs
\citep{fuj13e,ham14a}, which could power the jets. While the cold gas
accretion is generally more efficient in producing the AGN radiation,
the associated accretion processes could involve complicated physics,
such as dynamical instabilities and disc cooling. Thus, it is not
certain whether the AGN can respond sufficiently quickly and accordingly
to the changes in the environment. Finally, no correlation between AGN
jet power and total molecular gas mass is known to exist \citep{mcn11a},
which suggests that simple accretion of the molecular gas does not
control the AGN activities in BCGs at low redshifts.

In this paper, we investigate whether accretion onto the SMBHs in
BCGs can be characterised by the Bondi accretion in the broad sense, by
studying the correlation between the accretion rate and the AGN jet
power. We also compare the power available through the Bondi accretion
with the X-ray AGN and cool core luminosities to study the radiation
efficiency of the accretion discs and suppression of the cooling
flows. For these studies, we devise a new method to estimate the Bondi
accretion rate. The rest of the paper is organised as follows. In
Section~\ref{sec:method}, we describe the details of our method to
calculate the Bondi accretion rate. The data used in the analysis are
presented in Section~\ref{sec:data}, and the results are provided in
Section~\ref{sec:results}. Section~\ref{sec:discuss} is devoted to
discussion of the SMBH accretion and the heating of cluster cool
cores. Our main conclusions have been summarised in
Section~\ref{sec:con}. We assume cosmological parameters of $\Omega_{\rm
m0}=0.3$, $\Omega_{\rm \Lambda 0}=0.7$, and $h=0.7$, which are often
used in this field.  Unless otherwise noted, errors are the 1~$\sigma$
values.

\section{Method}
\label{sec:method}

We assume that hot gas accretion onto the SMBHs at the centres of our
sample galaxies is not dominated by the angular momentum and, therefore,
take it to be spherically-symmetric. The Bondi accretion rate is given
by
\begin{equation}
\label{eq:dotMB}
 \dot{M}_{\rm B} = 4\pi\lambda_{\rm c} 
(G M_\bullet)^2 c_{\rm s,B}^{-3}\rho_{\rm B}
= \pi\lambda_{\rm c} c_{\rm s,B} \rho_{\rm B} r_{\rm B}^2\:,
\end{equation}
where $M_\bullet$ is the SMBH mass, and $\rho_{\rm B}=\rho(r_{\rm B})$
and $c_{\rm s,B}=c_{\rm s}(r_{\rm B})$ are the density and the sound
speed at the Bondi accretion radius $r_{\rm B} = 2 G M_\bullet/c_{\rm
s,B}^2$ \citep{bon52a}. The coefficient $\lambda_{\rm c}$ depends on the
adiabatic index of the accreting gas ($\gamma$) and we assume
$\gamma=5/3$ and $\lambda_{\rm c}=0.25$. The sound speed $c_{\rm
s}=\sqrt{\gamma k_{\rm B} T/(\mu m_{\rm p})}$ is the function of gas
temperature $T$, where $\mu (=0.6)$ is the mean molecular weight, and
$m_{\rm p}$ is the proton mass. Equation (\ref{eq:dotMB}) shows that the
information on $M_\bullet$, $\rho_{\rm B}$, and $T_{\rm B} (=T(r_{\rm
B}))$ are required to evaluate $\dot{M}_{\rm B}$. Even with the superb
angular resolution of {\it Chandra}, an expected Bondi radius of any
system in our sample cannot be resolved. Thus, we need to extrapolate
the density and temperature at the innermost measurement radius ($r_{\rm
obs}$) to those at the Bondi radius. In previous studies, the
extrapolation was made by assuming a power-law for the density profiles
and a constant temperature \citep{all06a,bal08a}. However, it is not
certain whether such an assumption is physically justified. Moreover,
since the distances to the BCGs are generally large, this method could
cause large errors. Thus, a more physical method for the extrapolation
is required.

For this extrapolation, we made two assumptions. First, we assume that
the hot gas outside the Bondi radius is in a nearly hydrostatic
equilibrium:
\begin{equation}
\label{eq:hydroeq}
 -\frac{dP}{dr} = \rho g\:,
\end{equation}
where $P(r)$ is the thermal gas pressure, and $g(r)$ is the
gravitational acceleration. For some nearby clusters
\citep[e.g. Perseus;][]{fab06a}, it has been revealed that the innermost
region is strongly disturbed by AGN activities. However, the assumption
of the hydrostatic equilibrium can be still acceptable as long as the 
turbulent and
ram pressures associated with the gas motions are smaller than its thermal
pressure. While the actual gas velocity has not been measured, it will
be obtained with {\it Astro-H} in the near future. The second assumption
is that the gas temperature near the SMBH (i.e., at $r\sim r_{\rm B}$)
reflects the velocity dispersion $\sigma$ or the virial temperature
$T_{\rm gal,vir}$ of the host galaxy:
\begin{equation}
\label{eq:T0}
 T_0 = \beta^{-1} \frac{\mu m_{\rm p} \sigma^2}{k}\sim T_{\rm gal,vir}\:,
\end{equation}
where $k$ is the Boltzmann constant, and $\beta$ is the constant of
order of unity. Following \citet{mat01a}, we adopt $\beta=0.5$ for
massive elliptical galaxies including BCGs. The
second assumption is related to the first one, because the left-hand
side of equation~(\ref{eq:hydroeq}) is approximated by $-dP/dr\sim P/r=n
k T/r$, where $n$ is the number density of the gas, while the right-hand
side is approximated by
\begin{equation}
 \rho g = \rho\frac{G M(<r)}{r^2}\sim n\frac{kT_{\rm gal,vir}}{r}\:,
\end{equation}
where $M(<r)$ is the gravitational mass within the radius $r$. The
second assumption, i.e., equation~(\ref{eq:T0}), is generally consistent
with {\it ROSAT} X-ray observations \citep{mat01a}.  In
Section~\ref{sec:results}, we discuss the temperature profiles in the
central regions of nearby BCGs obtained with recent {\it Chandra}
observations.

We assume that the temperature profile reflects the size of the galaxy
and the profile between $r=r_{\rm B}$ and $r_{\rm obs}$ is given by
\begin{equation}
\label{eq:T}
 T(r) = T_0 + (T_{\rm obs} - T_0)\frac{\tanh(r/R_{\rm e})}{\tanh(r_{\rm
  obs}/R_{\rm e})}\:,
\end{equation}
where $T_{\rm obs}=T(r_{\rm obs})$ and $R_{\rm e}$ is the effective
radius (half-light radius) of the galaxy.  Thus, the temperature
decreases toward the galaxy centre in general. Once $T(r)$ is
determined, the Bondi radius $r_{\rm B}$ can be derived by numerically
solving the equation
\begin{equation}
\label{eq:rB}
 r_{\rm B} = \frac{2 G M_\bullet}{c_{\rm s}(T(r_{\rm B}))^2}
\end{equation}
for a given $M_\bullet$.

The equation of the hydrostatic equilibrium 
(equation~\ref{eq:hydroeq}) can be written as
\begin{equation}
\label{eq:drhodr}
 \frac{d\rho}{dr} = -\frac{\rho}{T}\left(\frac{\mu m_{\rm p}}{k}g 
+ \frac{dT}{dr}\right) \:.
\end{equation}
Since $T(r)$ has been determined by equation~(\ref{eq:T}), $\rho(r)$ can
be obtained by numerically integrating equation~(\ref{eq:drhodr}) and
setting $\rho_{\rm obs}=\rho(r_{\rm obs})$ and $g(r)$.  The electron
number density is given by $n_{\rm e}=\rho/(1.13\: m_{\rm p})$.

The gravitational acceleration $g$ is given by three components, i.e.,
$g=g_\bullet + g_{\rm gal} + g_{\rm cl}$, where $g_\bullet$ is the SMBH
contribution, $g_{\rm gal}$ is the galaxy contribution, and $g_{\rm cl}$
comes from the cluster \citep{mat06a,guo14a}. Thus,
\begin{equation}
\label{eq:gBH}
 g_\bullet = \frac{G M_\bullet}{r^2}\:.
\end{equation}
The acceleration from a galaxy with the Hernquist profile \citep{her90a}
is
\begin{equation}
 g_{\rm gal} = \frac{G M_{\rm gal}}{(r + r_{\rm H})^2}\:,
\end{equation}
where $M_{\rm gal}$ is the stellar mass of the galaxy, and $r_{\rm
H}=R_{\rm e}/1.815$. The cluster acceleration for the NFW profile
\citep{nav96a} is
\begin{equation}
 g_{\rm cl} = \frac{G M_{\rm vir}}{r^2}
\frac{\log(1+y)-y/(1+y)}
{\log(1+c_{\rm vir})-c_{\rm vir}/(1+c_{\rm vir})}\:,
\end{equation}
where $y=c_{\rm vir} r/r_{\rm vir}$, and $c_{\rm vir}$ is the
concentration parameter. The cluster virial radius, $r_{\rm vir}$,
is defined as the radius at which the average cluster density is
$\Delta(z)$ times the critical density $\rho_{\rm crit}(z)$ at the
cluster redshift $z$:
\begin{equation}
\label{eq:rvir}
 r_{\rm vir}=\left(\frac{3 M_{\rm vir}}{4\pi 
\Delta(z)\rho_{\rm crit}(z)}\right)^{1/3}\:.
\end{equation}
For $\Delta(z)$, we use the fitting formula of \citet{bry98a}:
$\Delta=18\pi^2 + 82x -39x^2$, where $x=\Omega_{\rm m}(z)-1$.

To summarise, we require the parameters $z$, $M_\bullet$, $M_{\rm
gal}$, $R_{\rm e}$, $\sigma$, $c_{\rm vir}$, and $M_{\rm vir}$, and the
boundary conditions $r_{\rm obs}$, $\rho_{\rm obs}$, and $T_{\rm obs}$
in order to obtain $r_{\rm B}$ and $\dot{M}_{\rm B}$. First, $T(r)$ is
determined by equations~(\ref{eq:T0}) and~(\ref{eq:T}) for given
$\sigma$, $T_{\rm obs}$, $R_{\rm e}$, and $r_{\rm obs}$. The Bondi
radius $r_{\rm B}$ is estimated by solving equation~(\ref{eq:rB}) for
given $T(r)$ and $M_\bullet$. Then, $\rho_{\rm B}=\rho(r_{\rm B})$ is
obtained by integrating equation~(\ref{eq:drhodr}) from $r=r_{\rm obs}$
to $r_{\rm B}$ using equations~(\ref{eq:gBH})--(\ref{eq:rvir}) for given
$T(r)$, $M_\bullet$, $M_{\rm gal}$, $R_{\rm e}$, $M_{\rm vir}$, $c_{\rm
vir}$ and $z$.  The Bondi accretion rate is given by
equation~(\ref{eq:dotMB}).

\begin{figure*}
  \begin{center}
    \begin{tabular}{c}
      \begin{minipage}{0.5\hsize}
        \begin{center}
          \includegraphics[width=84mm]{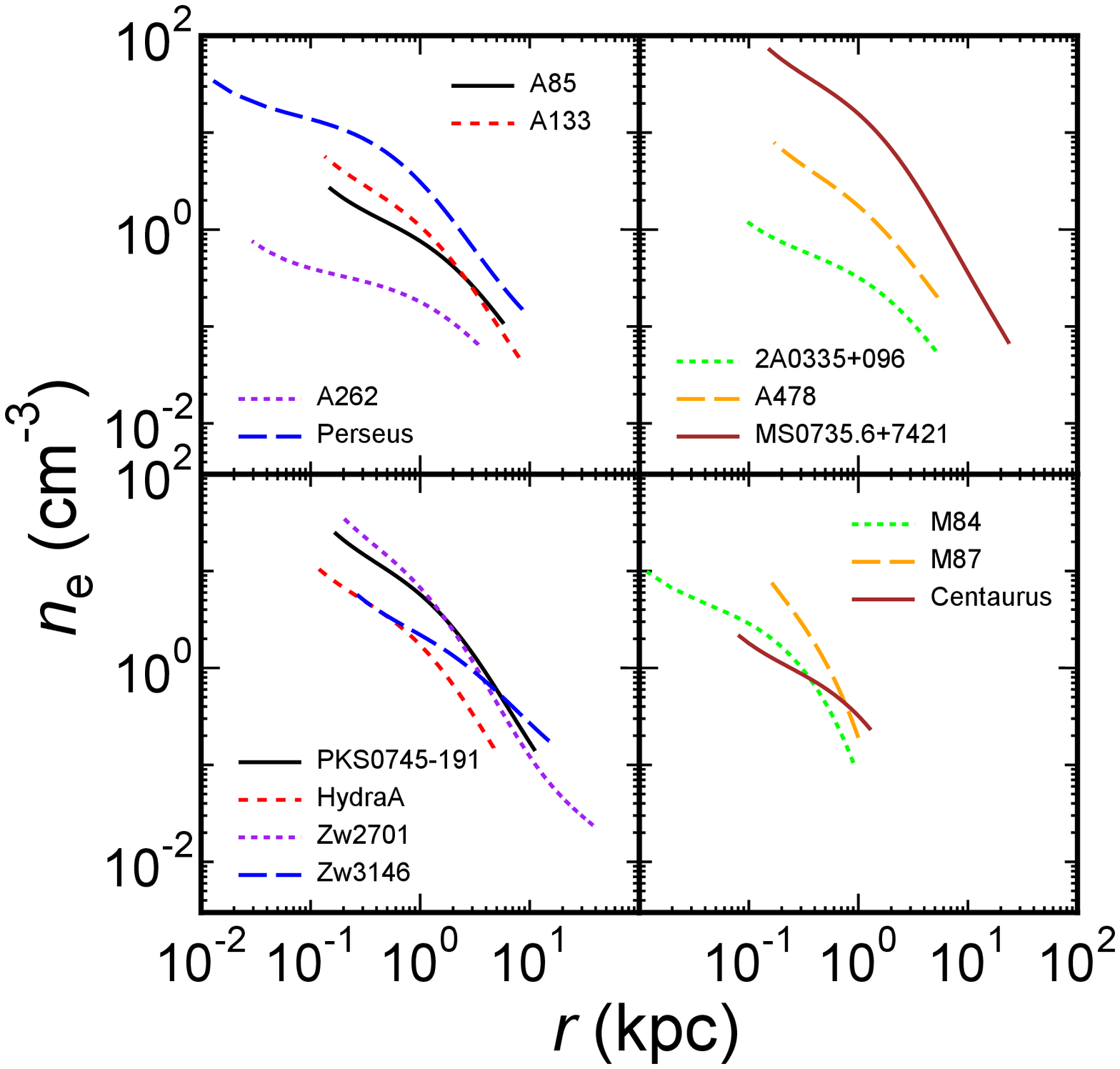}
        \end{center}
      \end{minipage}
      \begin{minipage}{0.5\hsize}
        \begin{center}
          \includegraphics[width=84mm]{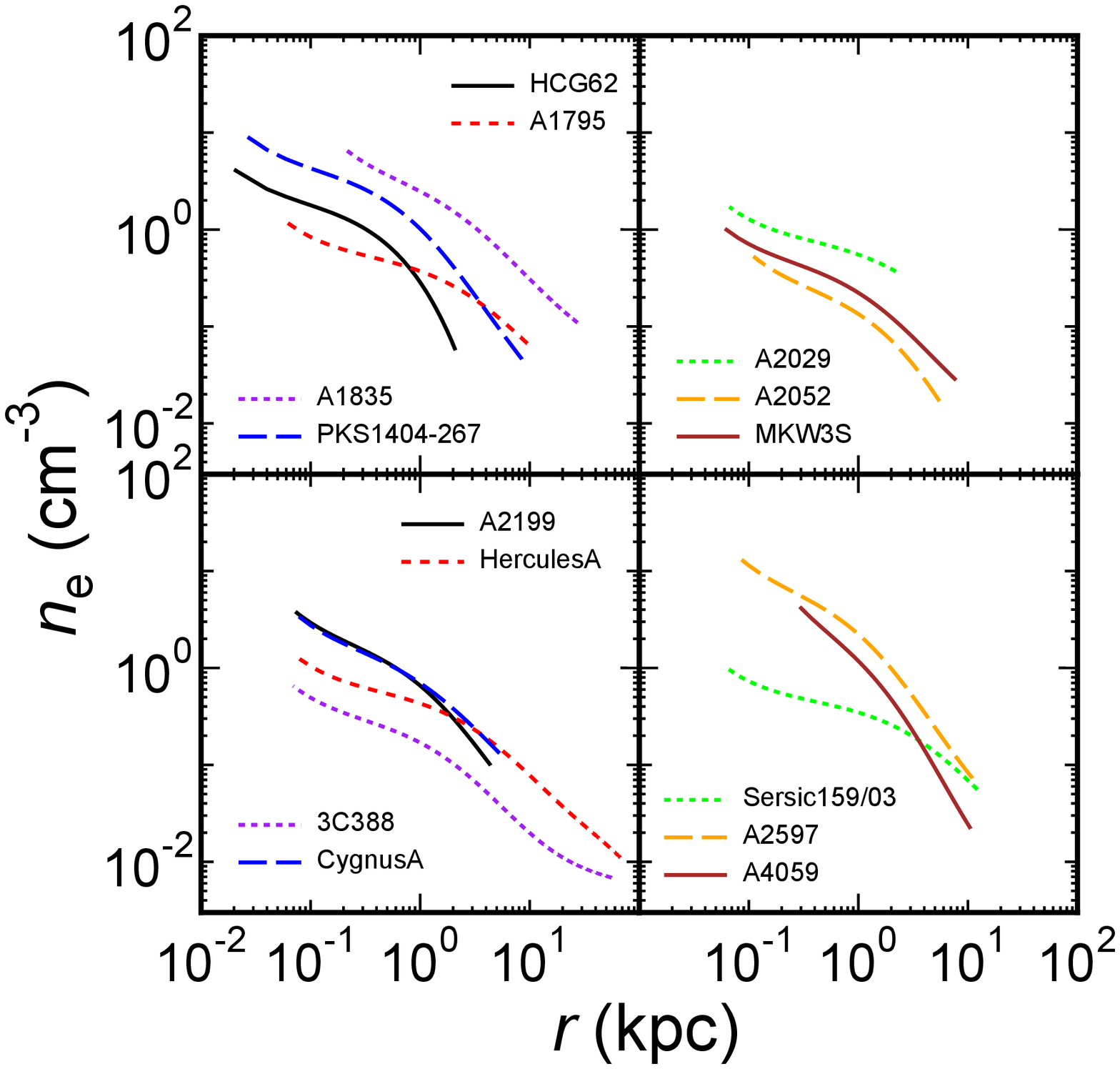}
        \end{center}
      \end{minipage}
    \end{tabular}
     \begin{tabular}{c}
      \begin{minipage}{0.5\hsize}
        \begin{center}
          \includegraphics[width=84mm]{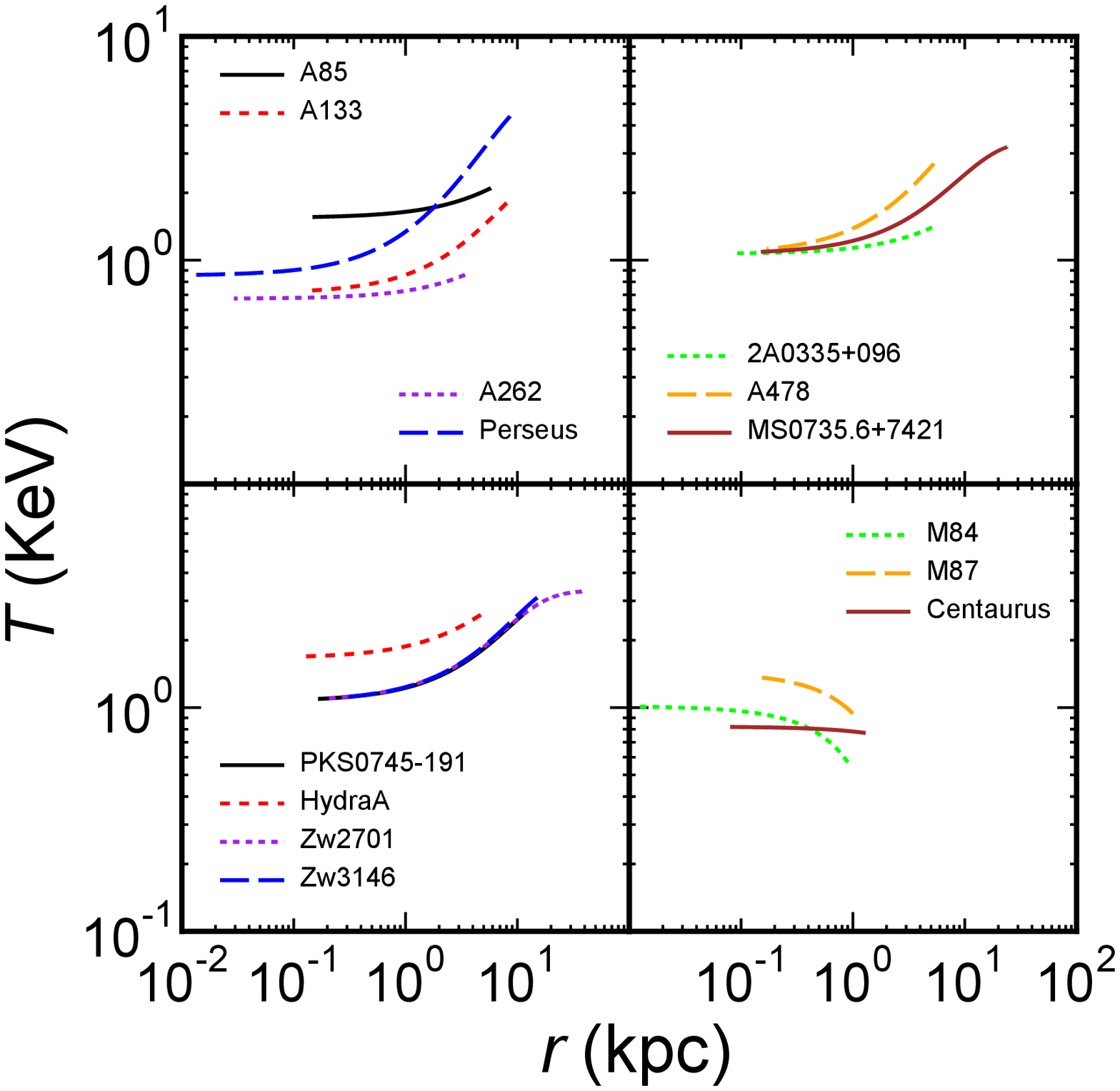}
        \end{center}
      \end{minipage}
      \begin{minipage}{0.5\hsize}
        \begin{center}
          \includegraphics[width=84mm]{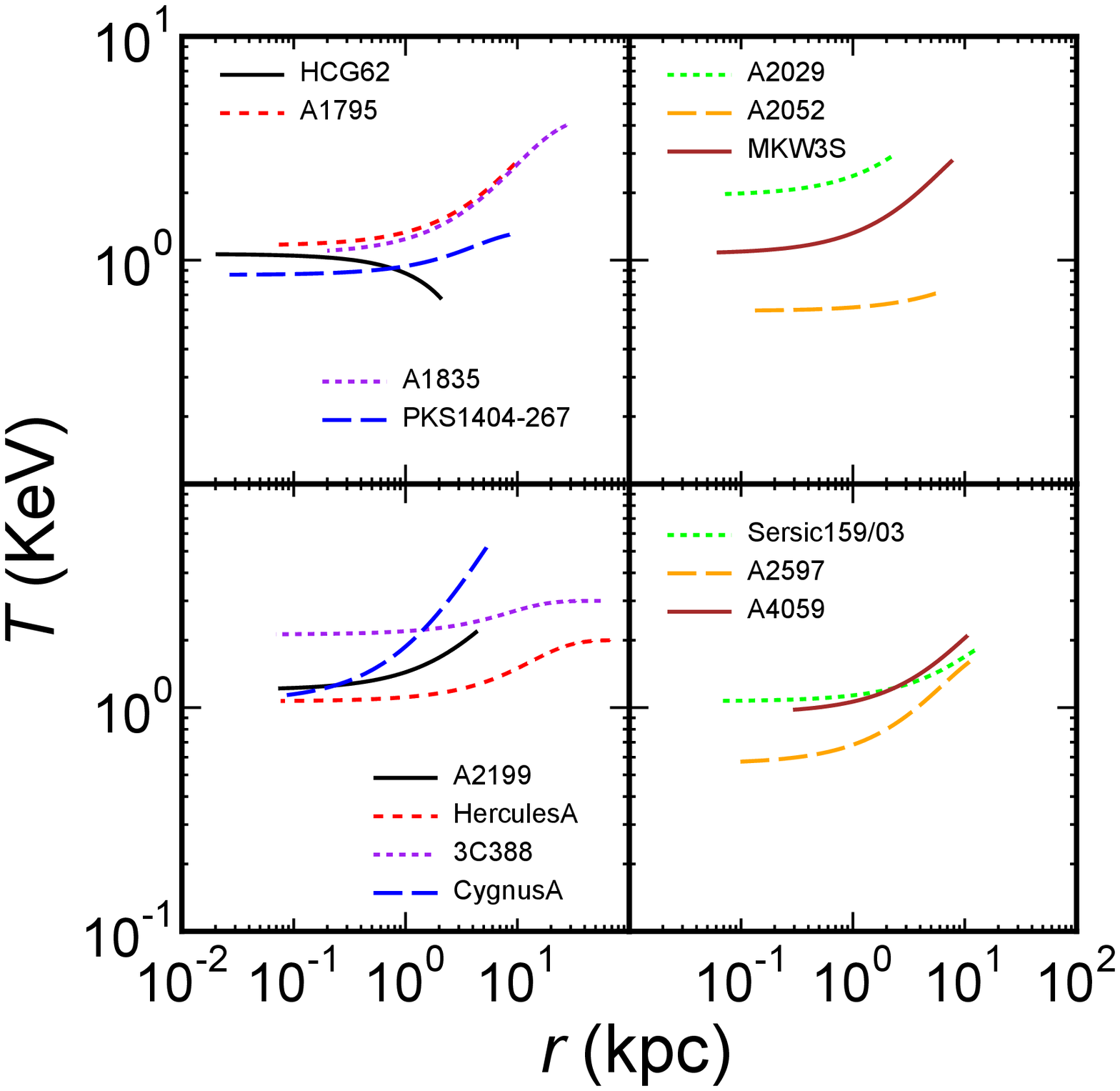}
        \end{center}
      \end{minipage}
    \end{tabular}
    \caption{Most probable density and temperature profiles for our
    sample galaxies. The right end left ends of each curve correspond to
    $r_{\rm obs}$ and $r_{\rm B}$, respectively.}  \label{fig:nT}
  \end{center}
\end{figure*}

\section{Data}
\label{sec:data}

Accounting for data uniformity and consistency, we study 28 bright
elliptical galaxies in clusters, for which the properties of the central
gas, the excavated cavity, etc., are well studied by \citet[][ see their
Table~6]{raf06a}. They are BCGs except for M84. The parameters for the
gravitational potentials are shown in Table~\ref{tab:pot}. For the
masses of the SMBHs, we adopt those derived by \citet{mcn11a}. The
masses were estimated using $R$-band absolute magnitudes ($M_{\rm R}$)
obtained by \citet{raf06a}. Since the errors in $M_\bullet$ were not
given, we take them as 0.5~dex, based on the dispersion around the
observed $M_{\rm R}$--$M_\bullet$ relations \citep[e.g.][]{mcl02a}. The
galaxy masses ($M_{\rm gal}$) were estimated by \citet{raf06a} using the
$R$-band absolute magnitudes, and thus they are consistent with
$M_\bullet$.  The effective radii of the galaxies ($R_{\rm e}$) are from
the 2MASS All-Sky Extended Source Catalogue\footnote{http:\slash\slash
irsa.ipac.caltech.edu\slash cgi-bin\slash Gator\slash nph-query}. We
take the average of $R_{\rm e}$ in the $J$, $H$ and $K$-bands, and their
scatter as the error. The velocity dispersions of the galaxies
($\sigma$) are taken from the HyperLeda
database\footnote{http:\slash\slash\ leda.univ-lyon1.fr\slash}. However,
12 galaxies have no data. For those galaxies, we use the error-weighted
average of the remaining 16 galaxies ($289.5\rm\: km\: s^{-1}$) as
$\sigma$, and the scatter of these galaxies, $\sim 289.5\rm\: km\:
s^{-1}$, as the error of $\sigma$ ($35.9\rm\: km\: s^{-1}$).

The parameters for the clusters are also shown in
Table~\ref{tab:pot}. Most of them are based on recent X-ray
observations. For the clusters with no appropriate X-ray data, we adopt
the data obtained through lensing observations or kinematics of the
member galaxies. For M84, HCG~62, and 3C~388, we do not consider the
contribution of the cluster component to the total gravitational
acceleration $g$, because M84 is not a BCG and there are no appropriate
data for the other two. Since the errors of $c_{\rm vir}$ and $M_{\rm
vir}$ were not given for MS~0735.6$+$7421 \citep{mol99a}, they are
assumed to be 0.3~dex. For Hercules~A and Cygnus~A, we use the cluster
temperatures and the core radii obtained by \citet{giz04a} and
\citet{smi02a}, respectively. The temperatures are converted into the
cluster masses $M_{\rm vir}$ by using the cluster mass--temperature
relation derived by \citet{sun09a}. The core radii ($r_{\rm c}$) are
converted into the characteristic radii ($r_{\rm s}=r_{\rm vir}/c_{\rm
vir}$) by using the relation of $r_{\rm s}=r_{\rm c}/0.22$
\citep{mak98a}.

The boundary conditions $r_{\rm obs}$, $n_{\rm e,obs} (=n_{\rm e}(r_{\rm
obs}))$, and $T_{\rm obs}$ shown in Table~\ref{tab:obs} are the same as
those in Table~6 of \citet{raf06a}. In their table, $r_{\rm obs}$,
$n_{\rm e,obs}$, and $T_{\rm obs}$ are represented by $a$, $n_{\rm e}$,
and $kT$, respectively. Although \citet{raf06a} give the average
densities and temperatures for $r<r_{\rm obs}$ excluding the AGN, most
of the emission comes from $r\sim r_{\rm obs}$, because the density
profiles near the galaxy centres are not very steep ($\alpha\la 1$ for
$\rho\propto r^{-\alpha}$) as is shown later. In other words, the
density profiles we have obtained do not produce excessively bright
emission from the gas in the vicinity of the AGN, and appear to be
consistent with the observations.

\section{Results}
\label{sec:results}

In Table~\ref{tab:res}, we present the Bondi accretion radii, $r_{\rm
B}$, the density, $n_{\rm e,B}=n_{\rm e}(r_{\rm B})$, and the
temperature, $T_{\rm B}=T(r_{\rm B})$, at these radii. The Bondi
accretion rates are also shown. The obtained Bondi radii are
substantially larger than those in \citet{raf06a}, because we adopt
smaller $T_{\rm B}$. The uncertainties of the results were estimated
using Monte Carlo simulations. We randomly perturbed each input
parameter with a Gaussian distribution of the perturbations, which had
an amplitude determined by the error bar of the parameter. We obtained
$10^4$ different realisations. Fig.~\ref{fig:nT} shows the density and
the temperature profiles between $r_{\rm B}$ and $r_{\rm obs}$. While
the density profiles can be represented by a power-law for most
clusters, some profiles show noticeable bends. This means that a
power-law is not always a good assumption when extrapolating the density
profile from $r_{\rm obs}$ to $r_{\rm B}$. For nearby galaxies such as
M84, M87, and Centaurus, the gas properties near the centres are
observationally known ($r_{\rm obs}\sim 1$~kpc). For these galaxies, the
temperatures at $r\sim r_{\rm obs}$ are close to $T_0$, which supports
our assumption that the gas in the central region of a galaxy is close
to the virial temperature of the galaxy.

The maximal power released from the neighbourhood of the SMBH through
the Bondi accretion is
\begin{equation}
 P_{\rm B} = \eta\dot{M}_{\rm B}c^2\:,
\end{equation}
where $\eta$ is the accretion efficiency assumed $\eta=0.1$. We compare
the Bond accretion power with the kinetic power of the jets, $P_{\rm
jet}$. We use the jet power estimated as the ratio of the enthalpy of
cluster X-ray cavities to their buoyancy timescales
\citep{raf06a,mcn11a}. The enthalpy is given by
\begin{equation}
\label{eq:Ecav}
 E_{\rm cav}=\frac{\gamma_{\rm c}}{\gamma_{\rm c}-1}P_{\rm s}V_{\rm c}\:,
\end{equation}
where $P_{\rm s}$ is the pressure of the gas surrounding the cavity,
$V_{\rm c}$ is the cavity's volume, and $\gamma_{\rm c}$ is the
adiabatic index of the gas inside the cavity. \citet{raf06a} and
\citet{mcn11a} assumed that the cavities are filled with
ultra-relativistic cosmic rays, which means that $\gamma_{\rm c}=4/3$
and $E_{\rm cav}=4\: P_{\rm s}V_{\rm c}$. However, it was recently
indicated that the cavities could be filled with low-energy cosmic rays
from the spectra of radio mini-halos \citep{fuj12a,fuj13a}. In that
case, $\gamma_{\rm c}$ is close to 5/3 and $E_{\rm cav}=2.5\: P_{\rm
s}V_{\rm c}$. Thus, we multiply $P_{\rm jet}$ in \citet{mcn11a} by
2.5/4, although the results are not much affected by this
modification. Note that while equation~(\ref{eq:Ecav}) is appropriate
for FR~I objects (most of our sample galaxies), it may underestimate the
jet power for FR~II objects (Cygnus~A) at most a factor of 10
\citep{ito08a}. Thus, $P_{\rm jet}$ for Cygnus~A should be regarded as a
lower-limit.

\begin{figure}
\includegraphics[width=84mm]{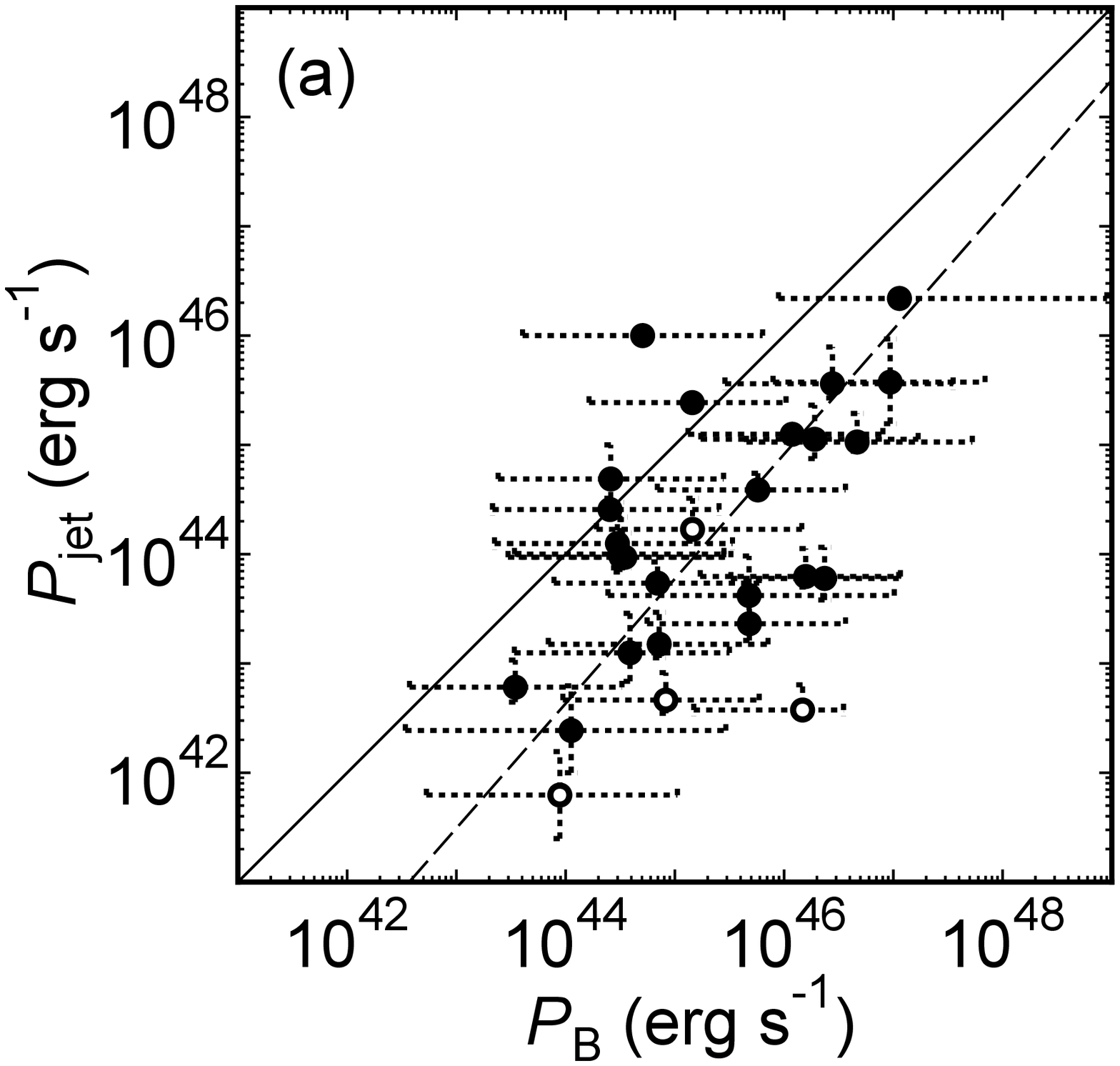}
\includegraphics[width=84mm]{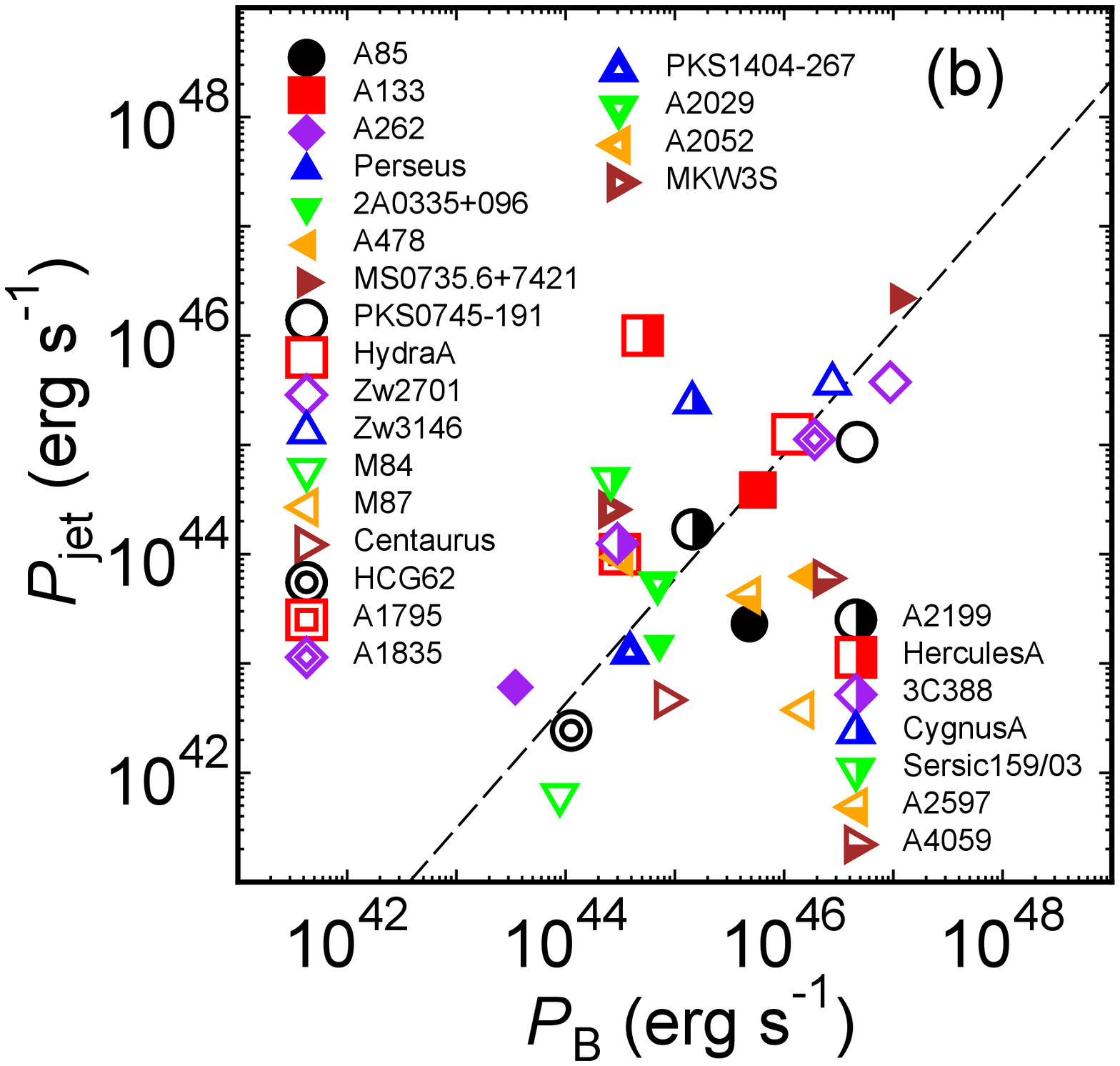} \caption{(a) Relation between
Bondi power $P_{\rm B}$ and jet power $P_{\rm jet}$. The dashed-line is
the best fitting model represented by equation~(\ref{eq:fit1}). The
solid line is $P_{\rm jet}=P_{\rm B}$.  The open circles are the
galaxies studied by \citet{all06a}. (b) Same as (a) but object names are
shown instead of error bars.}  \label{fig:pp}
\end{figure}

\begin{figure}
\includegraphics[width=84mm]{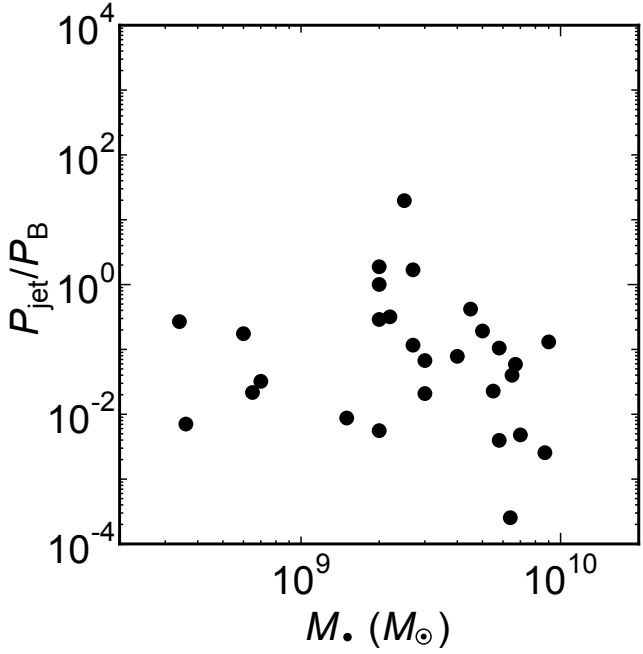} \caption{Relation between the mass
of SMBHs $M_\bullet$ and the ratio of the jet power $P_{\rm jet}$ to
the Bondi accretion power $P_{\rm B}$. Error bars are omitted for
clarity.}  \label{fig:Mpp}
\end{figure}

We present $P_{\rm jet}$ and $P_{\rm B}$ in Tables~\ref{tab:obs}
and~\ref{tab:res}, and display their relationship in
Fig.~\ref{fig:pp}. A positive correlation between $P_{\rm jet}$ and
$P_{\rm B}$ can be clearly seen. We found that the Spearman's rank
coefficient is 0.47. The probability that this is produced from a random
distribution is $P_{\rm null}=1.2\times 10^{-2}$. Using an ordinary
least-squares bisector regression method \citep{iso90a} and the
code\footnote{http:\slash\slash astrostatistics.psu.edu\slash
statcodes\slash sc\_regression.html}, the correlation can be described
as a power-law model of the form
\begin{equation}
\label{eq:fit1}
 \log\frac{P_{\rm jet}}{10^{42}\rm\: erg\: s^{-1}}
= A_1 + B_1\log\frac{P_{\rm B}}{10^{42}\rm\: erg\: s^{-1}}\:,
\end{equation}
where $A_1=-1.65_{-1.09}^{+1.57}$ and $B_1=1.14_{-0.22}^{+0.09}$. The
uncertainties were estimated by the Monte Carlo
simulations. Fig.~\ref{fig:Mpp} shows that $P_{\rm jet}/P_{\rm B}\la 1$
except for Hercules~A ($P_{\rm jet}/P_{\rm B}=20$), which means that the
Bondi accretion can power the jet activities in general, contrary to
previous studies \citep{mcn11a}. Note that $P_{\rm jet}/P_{\rm B}$ we
obtained are much smaller than those in Fig.~2 of \citet{mcn11a},
because they used the gas density and temperature at $r=r_{\rm obs}$ and
thus their resulting $\dot{M}_{\rm B}$ is generally smaller than ours.
Except for Hercules~A, the power ratios are in the range of $10^{-3}\la
P_{\rm jet}/P_{\rm B}\la 1$ and $\eta P_{\rm jet}/P_{\rm B}$ shows the
energy conversion efficiency from the rest mass energy of the infalling
gas to the jet power. The reason why Hercules~A exhibits a ratio that
high can be related to the ongoing merger detected by the {\it Hubble
Space Telescope} \citep{ode13a}, supplemented by injection of the cold
dusty gas in the region of $r_{\rm B}$ causing the hydrostatic
equilibrium to be strongly perturbed.

\section{Discussion}
\label{sec:discuss}

The correlation between $P_{\rm jet}$ and $P_{\rm B}$ shown in
Fig.~\ref{fig:pp} and the fact that $P_{\rm jet}/P_{\rm B}\la 1$, as
shown in Fig.~\ref{fig:Mpp}, indicate that the SMBH jet activities
of are controlled by accretion of the ambient hot gas. This leads to the
stable suppression of the cooling flows in cluster cores
(Section~\ref{sec:intro}). The correlation in Fig.~\ref{fig:pp} has a
large scatter. This may be partly related to the uncertainties of
observations. \citet{all06a} studied 9 nearby X-ray luminous elliptical
galaxies and found a tight $P_{\rm jet}$--$P_{\rm B}$
correlation. However, \citet{rus13b}, based on the revised analysis,
indicated that the correlation is much weaker. Our sample contains four
galaxies studied by \citet{all06a} --- M86, M87, NGC~4696 (Centaurus),
and NGC~6166 (A2199). The correlation among the four galaxies in
Fig.~\ref{fig:pp}a is not as tight as that in Fig.~4 of \citet{all06a}.
While \citet{all06a} studied the $P_{\rm jet}$--$P_{\rm B}$ relation
over 2.5 orders of magnitude in $P_{\rm jet}$, we studied the relation
over 4 orders of magnitude (Fig.~\ref{fig:pp}), which may be the reason
that we found the correlation in spite of the large scatter.

Uncertainties in our model may also be responsible for the scatter. In
the case of the Centaurus, for example, we predict that the gas density
continues to increase and the temperature is nearly constant
toward $r_{\rm B}$ (Fig.~\ref{fig:nT}). On the contrary, the latest {\it
Chandra} observations detected a flatter density profile and a slightly
rising temperature at $r\la 1$~kpc \citep{rus13b}, which may be due to
gas heating by the AGN. This heating may cause $\dot{M}_{\rm B}$ to
decrease at $r_{\rm B}$. For the Perseus cluster, {\it Chandra}
observations have shown that the temperature seems to bottom out at
$\sim 3$~keV instead of $T_0\sim 1$~keV, and that the central region
($\la 10$~kpc) is strongly disturbed by the cavities \citep{fab06a}.

\begin{figure}
\includegraphics[width=84mm]{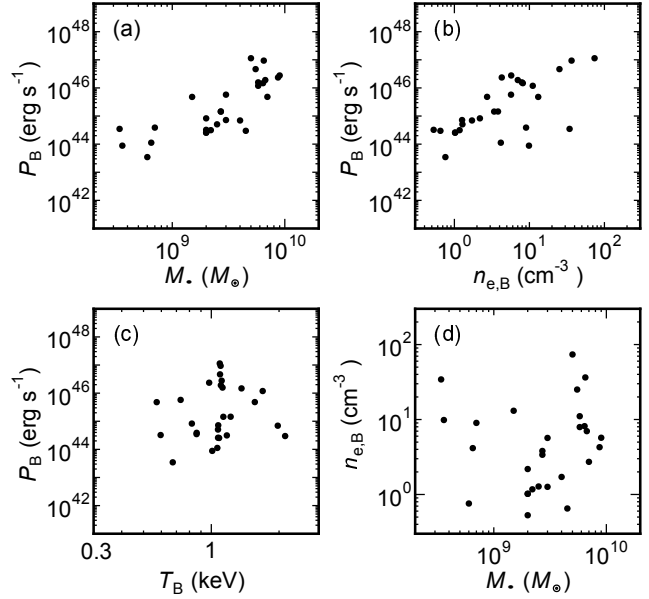} \caption{(a) Mass of the SMBH
$M_\bullet$ vs. Bondi accretion power $P_{\rm B}$. (b) Electron density
at the Bondi radius $n_{\rm e,B}$ vs. Bondi accretion power $P_{\rm
B}$. (c) Temperature at the Bondi radius $T_{\rm B}$ vs. Bondi accretion
power $P_{\rm B}$. (d) Mass of the SMBH $M_\bullet$ vs. electron
density at the Bondi radius $n_{\rm e,B}$. Error bars are omitted for
clarity.}  \label{fig:4dotM}
\end{figure}

Equation~(\ref{eq:dotMB}) shows that the Bondi accretion rate
$\dot{M}_{\rm B}$ is determined by $M_\bullet$, $\rho_{\rm B}$ and
$T_{\rm B}$.  The correlations between $M_\bullet$ and $P_{\rm B}$, and
between $n_{\rm e,B}$ and $P_{\rm B}$ are noticeable
(Figs.~\ref{fig:4dotM}a and~\ref{fig:4dotM}b). On the other hand, there
is no correlation between $T_{\rm B}$ and $P_{\rm B}$, and between
$M_\bullet$ and $n_{\rm e,B}$ (Figs.~\ref{fig:4dotM}c and
\ref{fig:4dotM}d). Therefore, the latter two parameters affect
$\dot{M}_{\rm B}$ independently. Since $n_{\rm e,B}$ depends on $n_{\rm
obs}$, the jet activity may be regulated by their past gas heating at
$r\ga r_{\rm obs}$. No detection of correlation between $T_{\rm B}$ and
$P_{\rm B}$ (Fig.~\ref{fig:4dotM}c) results from the scatter of $T_0$
among our sample galaxies being small.

\begin{figure}
\includegraphics[width=84mm]{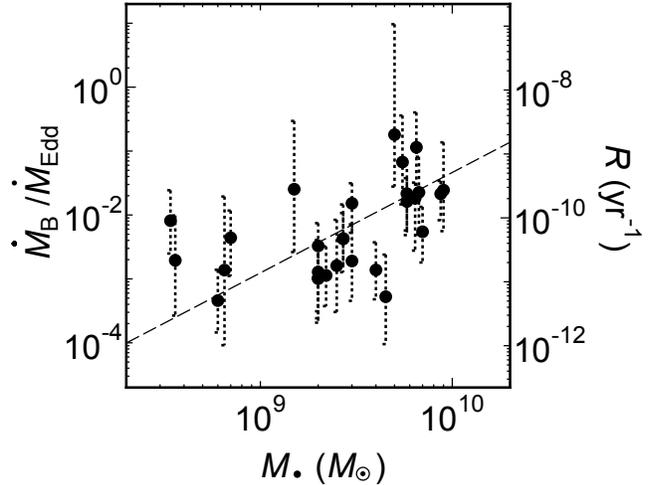} \caption{Mass of the SMBHs
$M_\bullet$ vs.  Eddington-scaled Bondi accretion rate $\dot{M}_{\rm
B}/\dot{M}_{\rm Edd}$ or specific growth rate ${\cal R}=\dot{M}_{\rm
B}/M_{\rm \bullet}$. The dashed-line is the best fitting model
represented by equation~(\ref{eq:fit2}).}  \label{fig:MEdd}
\end{figure}

\begin{figure}
\includegraphics[width=84mm]{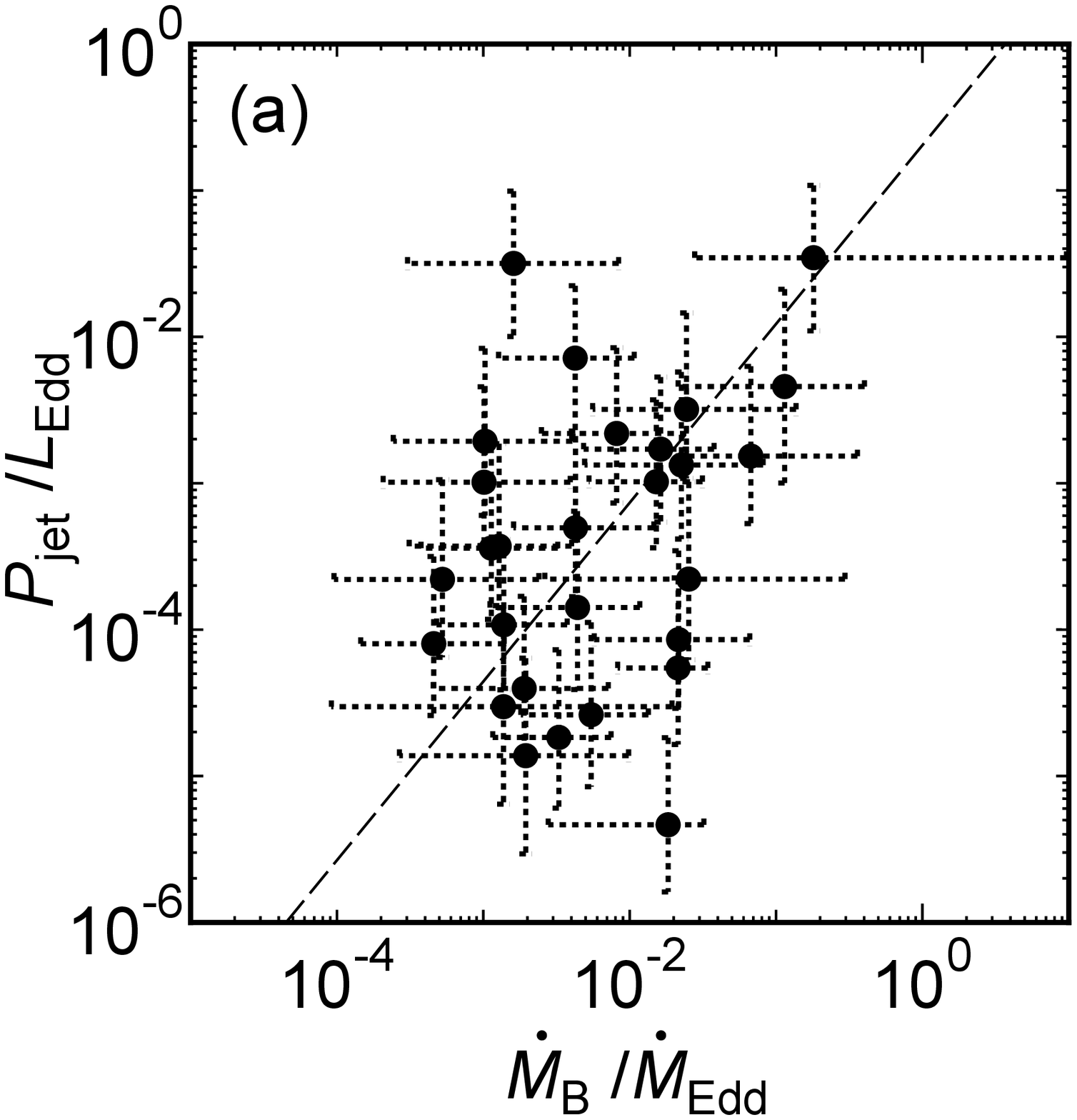}
\includegraphics[width=84mm]{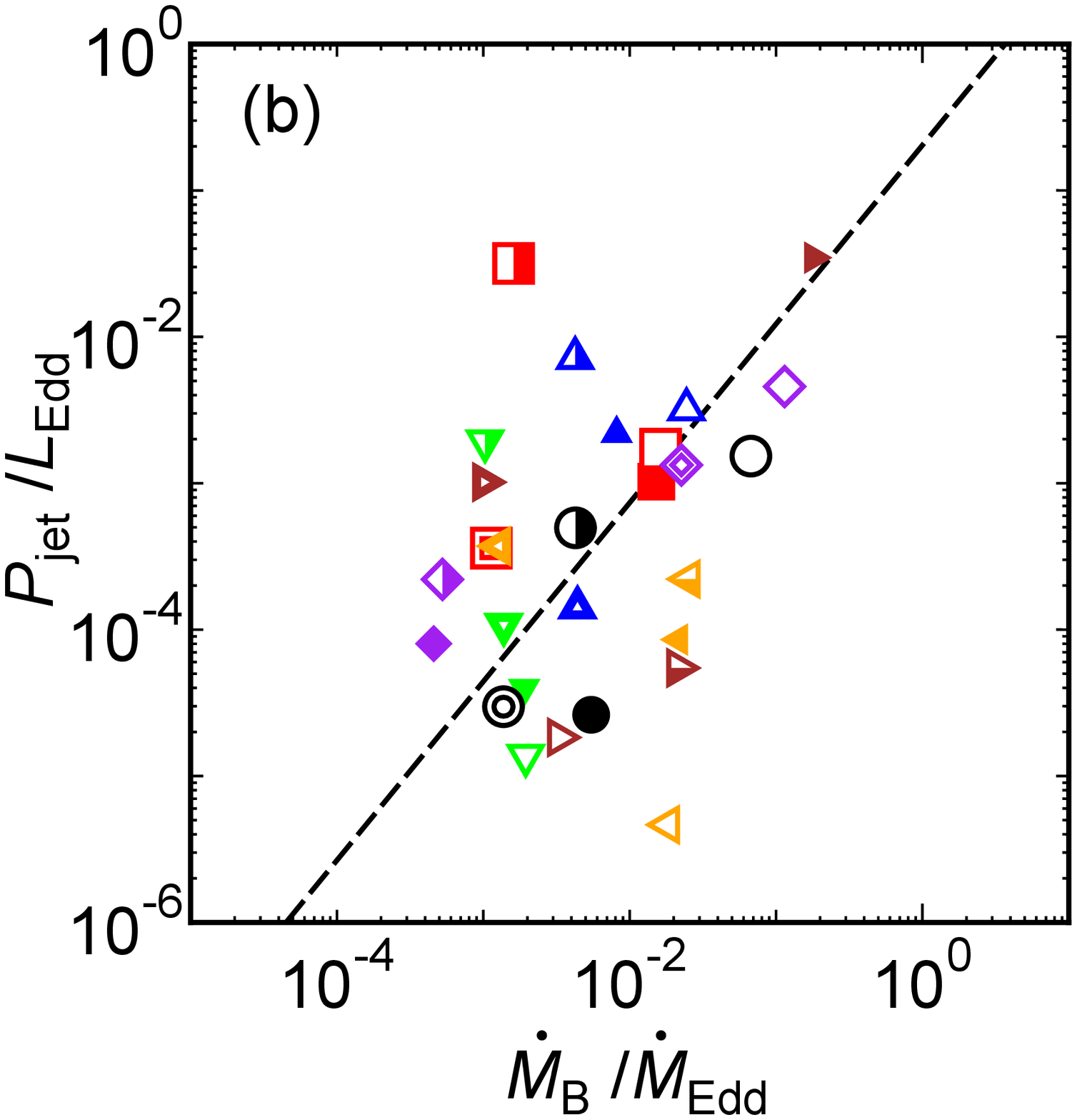}
    \caption{(a) Relation between Eddington-scaled Bondi accretion rate
   $M_{\rm B}/M_{\rm Edd}$ and Eddington-scaled jet power $P_{\rm
   jet}/L_{\rm Edd}$. The dashed-line is the best fitting model
   represented by equation~(\ref{eq:fit3}). (b) Same as (a) but object
   names are shown by the marks in Fig.~\ref{fig:pp}b.}  \label{fig:Mp}
\end{figure}

Studies have indicated that the properties of the accretion disc around
a SMBH depend on the gas accretion rate normalised by the Eddington
accretion rate \citep{nar08a,yua14a},
\begin{equation}
 \dot{M}_{\rm Edd}=L_{\rm Edd}/(\eta_r c^2)\:,
\end{equation}
where $\eta_r = 0.1$ is the radiation efficiency, and $L_{\rm Edd}$ is
the Eddington luminosity:
\begin{equation}
 L_{\rm Edd} = 1.26\times 10^{38}\: \left(\frac{M_\bullet}{M_\odot}
\right)\rm\: erg\: s^{-1}\:
\end{equation}
\citep[e.g.][]{nar08a}. In Fig.~\ref{fig:MEdd}, we show a plot of
$M_\bullet$ versus $\dot{M}_{\rm B}/\dot{M}_{\rm Edd}$. A positive
correlation is seen and the best fitting relation is
\begin{equation}
\label{eq:fit2}
 \log\left(\frac{\dot{M}_{\rm
B}}{\dot{M}_{\rm Edd}}\right)
= A_2 + B_2\log\left(\frac{M_\bullet}{10^9\: M_\odot}\right)\:,
\end{equation}
where $A_2=-2.91_{-0.47}^{+0.57}$ and $B_2=1.57_{-0.18}^{+0.45}$. The
probability that this is produced from a random distribution is only
$P_{\rm null}=1.8\times 10^{-3}$. The positive correlation reflects the
relation between $M_\bullet$ and $P_{\rm B}$ shown in
Fig.~\ref{fig:4dotM}a. This is in contrast with the trend for
$\sim$\,23,000 type~2 AGNs in the Sloan Digital Sky Survey (SDSS).
\citet{hec04a} indicated that most present-day accretion occurs onto
SMBHs with masses less than $10^8\: M_\odot$, and the accretion onto
more massive SMBHs is inefficient and, therefore, substantially
sub-Eddington.

A simple explanation can be that the hot accretion flows in BCGs are
maintained by continuous heating from the AGN.  The gas accretion rate
normalised by the Eddington accretion rate $\dot{M}_{\rm B}/M_{\rm Edd}$
is proportional to the black hole specific growth rate ${\cal
R}=\dot{M}_{\rm B}/M_\bullet$. Fig.~\ref{fig:MEdd} shows that the growth
timescale (${\cal R}^{-1}$) is smaller than the age of the galaxies
($\sim 10$~Gyr) for $M_\bullet\ga 3\times 10^9\rm\: M_\odot$, and the
specific growth rate for the most massive SMBHs ($\sim 10^{10}\rm\:
M_\odot$) is ${\cal R}\sim 0.5\:{\rm Gyr}^{-1}$. Although it is not
certain whether the high accretion rates are maintained for a long time,
they might have caused rapid growth of those SMBHs. This accretion
may lead to the formation of SMBHs with $M_\bullet\sim 10^{10}\:
M_\odot$.

Fig.~\ref{fig:Mp} shows the relation between $\dot{M}_{\rm
B}/\dot{M}_{\rm Edd}$ and $P_{\rm jet}/L_{\rm Edd}$. The best fitting
relation is
\begin{equation}
\label{eq:fit3}
 \log\left(\frac{P_{\rm jet}}{L_{\rm Edd}}\right)
= A_3 + B_3\log\left(\frac{\dot{M}_{\rm
B}}{\dot{M}_{\rm Edd}}\right)\:,
\end{equation}
where $A_3=-0.69_{-1.18}^{+0.76}$ and $B_3=1.22_{-0.25}^{+0.04}$.
However, the null probability is relatively large, i.e., $P_{\rm
null}=0.15$. Thus, Fig~\ref{fig:Mp} displays that $P_{\rm
jet}/L_{\rm Edd}$ has a large scatter for a given $\dot{M}_{\rm
B}/\dot{M}_{\rm Edd}$.  The poor correlation among the above scaled
values means that the relation between $P_{\rm B}$ and $P_{\rm jet}$ in
Fig.~\ref{fig:pp} does not depend on the black hole mass alone (see also
Fig.~\ref{fig:Mpp}). Moreover, the underlying accretion disk
properties and the jet production efficiency do not depend only on
$\dot{M}_{\rm B}/\dot{M}_{\rm Edd}$.

\begin{figure}
\includegraphics[width=84mm]{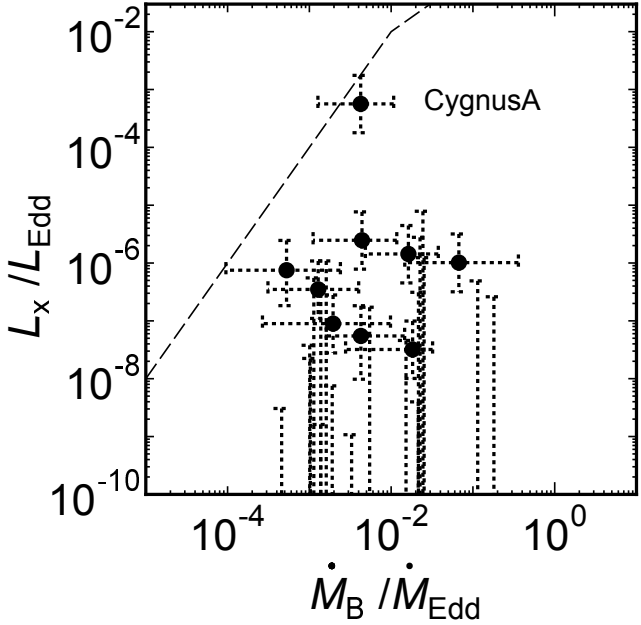} \caption{Eddington-scaled Bondi
accretion rate $\dot{M}_{\rm B}/\dot{M}_{\rm Edd}$ and Eddington-scaled
X-ray luminosity $L_{\rm X}/L_{\rm Edd}$. The dashed-line is a relation
for X-ray binaries represented by equations~(\ref{eq:bin1})
and~(\ref{eq:bin2}).}  \label{fig:MEddLX}
\end{figure}

We also found that the AGN luminosities do not necessarily increase with
the accretion rate ratio $\dot{M}_{\rm B}/\dot{M}_{\rm Edd}$. In
Fig.~\ref{fig:MEddLX}, we present a plot of $\dot{M}_{\rm B}/\dot{M_{\rm
Edd}}$ versus $L_{\rm X}/L_{\rm Edd}$, where $L_{\rm X}$ is the X-ray
luminosity derived from the X-ray flux obtained by \citet{rus13b},
except for Perseus (Table~\ref{tab:obs}). The X-ray flux from the AGN in
the Perseus cluster is too large to measure accurately \citep{rus13b},
and it is not included in the figure. Fig.~\ref{fig:MEddLX} shows that
the AGN are very dim and for most of them only upper limits have been
obtained. The dashed line is the relation represented by
\begin{equation}
\label{eq:bin1}
 \frac{L_{\rm X}}{L_{\rm Edd}} = \frac{\dot{M}_{\rm
B}}{\dot{M}_{\rm Edd}}
\end{equation}
for $\dot{M}_{\rm B}/\dot{M}_{\rm Edd} > 0.01$ and
\begin{equation}
\label{eq:bin2}
 \frac{L_{\rm X}}{L_{\rm Edd}} = 100 \left(\frac{\dot{M}_{\rm
B}}{\dot{M}_{\rm Edd}}\right)^2
\end{equation}
for $\dot{M}_{\rm B}/\dot{M}_{\rm Edd} < 0.01$, which is often applied
to stellar-mass black holes in the Galaxy \citep[e.g.][]{nar08a}.
Fig.~\ref{fig:MEddLX} shows that all the AGN in our sample are much
dimmer than that relation and they are extremely radiatively
inefficient, especially when compared to the X-ray binaries
\citep[e.g.][]{esi97a}. Although some of the non-detected sources could
be heavily absorbed, it is unlikely that the absorption alone can
explain the overall dimness of the AGN \citep{rus13b}.  The dimness of
the AGNs in elliptical galaxies has also been pointed out in previous
studies \citep{dim00a,loe01a,pel05a,bal08a}. Fig.~\ref{fig:MEddLX} shows
that some galaxies have accretion rates that are even larger than
$\dot{M}_{\rm B}/\dot{M}_{\rm Edd}=0.01$, above which the accretion
discs are expected to become radiatively efficient
\citep{nar08a,yua14a}. Their low X-ray luminosities shown in
Fig.~\ref{fig:MEddLX} may indicate that $\dot{M}_{\rm B}/\dot{M}_{\rm
Edd}$ is not the only parameter that determines the radiative efficiency
of accretion discs.  BCGs at low-redshifts are immersed in the hot
intracluster medium (ICM), which is being continuously heated by the AGN
(otherwise cooling flows should have developed). Within this framework,
the hot gas accretion may dominate and radiatively inefficient accretion
may be allowed even if $\dot{M}_{\rm B}/\dot{M}_{\rm Edd}\ga 0.01$.  The
statistical significance of the above effect, however, is not clear, as
only 3 galaxies (MS~0735.6+7421, PKS~0745-191, and Zw~2701) exceed
accretion rates $\dot{M}_{\rm B}/\dot{M}_{\rm Edd}\sim 0.01$ by more
than $1\sigma$.

\begin{figure}
\includegraphics[width=84mm]{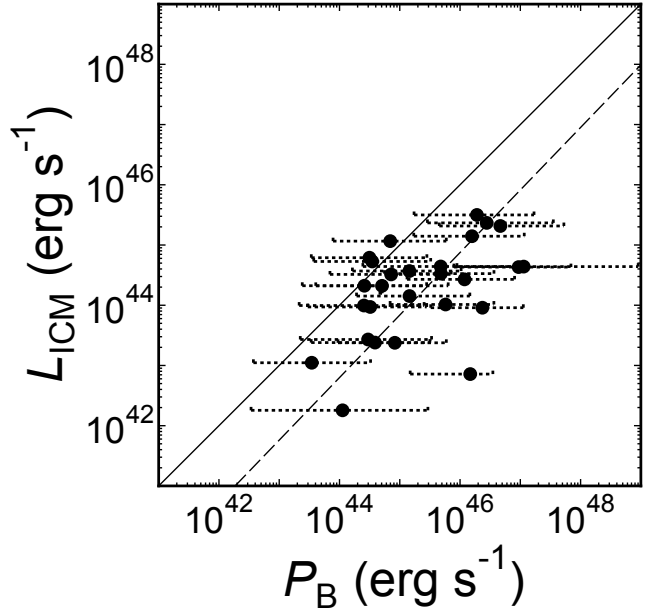} \caption{Relation between Bondi
power $P_{\rm B}$ and X-ray luminosity of the cool core $L_{\rm
ICM}$. The errors of $L_{\rm ICM}$ are comparable or smaller than the
sizes of the filled circles. The dashed-line is the best fitting model
represented by equation~(\ref{eq:fit4}). The solid line shows $L_{\rm
ICM}=P_{\rm B}$.}  \label{fig:PBLc}
\end{figure}

Fig.~\ref{fig:PBLc} shows the Bondi accretion power ($P_{\rm B}$) and
the X-ray luminosity of the ICM inside the cluster cooling radius that
is offset to be consistent with the spectra ($L_{\rm ICM}=L_{\rm
Xc}-L_{\rm cool}$). The luminosities are derived by \citet{raf06a};
$L_{\rm Xc}$ is the X-ray luminosity within which the gas has a cooling
time less than $7.7\times 10^9$~yr (the look-back time for $z=1$), and
$L_{\rm cool}$ is the associated (cooling) luminosity of gas cooling to
low temperatures, derived from the X-ray spectrum. Since $L_{\rm cool}$
could not be determined for A1835 \citep{raf06a}, we assume $L_{\rm
cool}=0$ for the cluster. However, since $L_{\rm cool}\ll L_{\rm Xc}$
for most other clusters, the assumption will not affect the results
strongly.

Fig.~\ref{fig:PBLc} indicates that $P_{\rm B}$ and $L_{\rm ICM}$
are tightly correlated. It suggests strongly that the radiative energy
loss of the cluster cool cores is being compensated by the AGN
feedback. The best fitting relation is
\begin{equation}
\label{eq:fit4}
 \log\left(\frac{L_{\rm ICM}}{10^{42}\rm\: erg\: s^{-1}}\right)
= A_4 + B_4\log\left(\frac{P_{\rm B}}
{10^{42}\rm\: erg\: s^{-1}}\right)\:,
\end{equation}
where $A_4=-1.28_{-0.97}^{+1.31}$ and $B_4=1.04_{-0.19}^{+0.07}$. The
null probability is $P_{\rm null}=3.1\times 10^{-3}$. The index $B_4$ is
consistent with unity, which means that $P_{\rm B}$ is proportional to
$L_{\rm ICM}$. The fact that $L_{\rm ICM}/P_{\rm B}\la 1$ means that
the Bondi accretion power is large enough to offset radiative cooling of
the cool cores. Since $L_{\rm ICM}/P_{\rm B}\sim 0.1$ on average and
$\eta=0.1$, about 1\% of the rest mass energy of the accreted gas that
passed $r_{\rm B}$ was used to heat up the cool cores. The
correlation described by equation~(\ref{eq:fit4}) may reflect
the fact that lower entropy (higher density and/or lower temperature) of
the gas leads to both larger $\dot{M}_{\rm B}$ (equation~\ref{eq:dotMB})
and larger $L_{\rm ICM}$. However, since $\dot{M}_{\rm B}$ is also
sensitive to $M_\bullet$ and $g(r)$ (equation~\ref{eq:drhodr}), the
correlation is not obvious.

\citet{fuj04c} showed that no correlation exists between $M_\bullet$ and
$L_{\rm ICM}$. This may indicate that the accretion rate fluctuates with
time. \citet{bir04a} and \citet{raf06a} compare $P_{\rm jet}$ with
$L_{\rm ICM}$ and found a correlation between them. However, for a
significant fraction of their sample clusters, they have shown that
$P_{\rm jet}<L_{\rm ICM}$, which may mean that the jet power alone
is not large enough to balance the cooling. Combining the relation
$P_{\rm jet}\la L_{\rm ICM}$ with our results $P_{\rm jet}\la P_{\rm B}$
(Fig.~\ref{fig:Mpp}) and $P_{\rm B}\ga L_{\rm ICM}$
(Fig.~\ref{fig:PBLc}), we obtain $P_{\rm jet}\la L_{\rm ICM}\la P_{\rm
B}$.  Assuming that $L_{\rm ICM}/P_{\rm B}$ of the Bondi power is used
to heat the cool core, a large portion of the Bondi power ($\sim (L_{\rm
ICM}-P_{\rm jet})/P_{\rm B}$) may be transferred to the ICM through the
form that does not appear in the jet power $P_{\rm jet}$ estimated from
the cavity volumes (equation~\ref{eq:Ecav}). This energy can be
transferred by the cosmic rays escaping from the cavities
\citep{guo08a,fuj13a}. However, since the AGN activities are
intermittent, time-averaged jet power $\bar{P}_{\rm jet}$ may be close
to the ICM luminosity $L_{\rm ICM}$.  It is also likely that not all
cavities have been observed, and that $P_{\rm jet}$ has been
underestimated.

The discussions so far are based on the Bondi accretion
\citep{bon52a}. However, whether the Bondi accretion is actually
realised in galaxies have been debated in the literature. For example,
small angular momentum of the accreting gas may significantly reduce the
accretion rate by introducing a centrifugal barrier
\citep{pro03a,kru05a}, although viscosity of the hot gas may transport
the angular momentum \citep{par09a,ino10b,nar11a}. Moreover, owing to
the high density and the short cooling time of the hot gas around the
SMBH, thermal instability may develop there and a substantial fraction
of the hot gas may turn into cold gas
\citep{sok06a,bar12a,sha12a,gas13a}. It is interesting to note that even
if this happens, the apparent X-ray temperature does not necessarily
decrease near the galactic centre, in spite of the efficient cooling
\citep{gas13a}. {\it Chandra} observations show that the hot gas density
drops while its temperature is almost constant at $\la 10$~kpc from the
centre of the Perseus cluster \citep{fab06a}. Although the density
profile might have been affected by the hot cavities or temporal AGN
heating, it could also be the result of the removal of the hot gas by
the thermal instability which triggers a cold gas flow in the
region. The angular momentum initially retained by turbulent hot gas may
decrease after the development of the thermal instability through
collisions between cold blobs \citep{shlo90}, which can also lead to the
formation of a turbulent cold disk and develop a large accretion rate at
the event horizon of the black hole $\dot{M}_{\rm BH}$
\citep{piz10a,gas14a}. Furthermore, outflows may significantly modify
the accretion flow and may make $\dot{M}_{\rm BH}$ much smaller than the
apparent Bondi accretion rate $\dot{M}_{\rm B}$
\citep{bla99a,yua12a}. If $\dot{M}_{\rm BH}\ll \dot{M}_{\rm B}$, other
energy sources such as black hole spins are required for efficient jet
production \citep{nem14a}. On the other hand, outflows can solve the
problem of low radiative efficiency of the SMBHs
(Fig.~\ref{fig:MEddLX}), because $L_{\rm X}$ should reflect
$\dot{M}_{\rm BH}$, which could be much smaller than $\dot{M}_{\rm B}$.

The correlations found in this study do not necessary mean that the
accretion flows in BCGs are represented by the `classical' and
oversimplified Bondi accretion flow. Significant departures from such a
flow, which involve angular momentum, multi-phase gas, and time- and
space-averaging should be expected. Hence the hot gas accretion flows
appear to reflect the Bondi model in some broad sense.  At least, it
reflects the density and temperature (or entropy) of the hot gas at
$r\sim r_{\rm obs}$ ($\sim 1$--70~kpc; Table~\ref{tab:obs}), because we
estimated $r_{\rm B}$ and $\dot{M}_B$ from them. For example, the
thermal instability turns on at some radius $r_{\rm inst}$ ($< r_{\rm
obs}$), and the emerging cold gas flow ($r\la r_{\rm inst}$) is
influenced by the boundary conditions at $r\sim r_{\rm obs}$.

\section{Conclusions}
\label{sec:con}

We have investigated the Bondi accretion onto the SMBHs in BCGs. For
that purpose, we devised a new method to estimate the Bondi accretion
rate even when the Bondi radius is under-resolved in the X-ray
observations. Our method is based on two assumptions: (1) the gas is
in nearly a hydrostatic equilibrium, and (2) the gas temperature is
represented by the virial temperature of the galaxy near its centre.

We have applied our method to 28 galaxies and have obtained the Bondi
accretion rates. We found a correlation between the Bondi accretion
power and the power of the jets associated with the SMBHs. We also found
that the jet power can be well supported by the Bondi accretion, in
contrast with previous studies.  These results indicate that the AGN
feedback in the BCGs is controlled by the accretion of the surrounding
hot gas, whose origin lies in a stable heating of the cluster cool
cores. The specific growth rates of the SMBHs increase as their mass
increases, which may explain the existence of hypermassive SMBHs ($\sim
10^{10}\: M_\odot$). The Eddington X-ray luminosities of the AGN are
very small compared to their Eddington luminosities, even if the
accretion rates are close to the Eddington accretion rate. This may
suggest that massive gas accretion with a low radiative efficiency is
realised in BCGs. Moreover, we have found that the Bondi accretion power
correlates linearly with the X-ray luminosities of the cluster cool
cores. We have shown that the power is large enough to offset the
radiative cooling of the cool core. These results may indicate that the
cooling of the cool cores is well balanced with the AGN feedback
associated with the Bondi accretion.

While the correlations we found do not necessarily mean that the
gas accretion on the SMBHs in BCGs follows the simplified Bondi 
prescription, they demonstrate that the accretion onto the SMBHs 
follows it in the broad sense, in some time- and space-averaged sense.

\section*{Acknowledgments}

We are grateful to the referee for valuable comments.  We thank
M.~Gaspari, Y.~Kunisawa, K.~Nagamine, N.~Soker, and F.~Yuan for useful
discussions. This work was supported by International Joint Research
Promotion Program by Osaka University. NK acknowledges the financial
support of Grant-in-Aid for Young Scientists
(B:25800099). I.S. acknowledges partial support from the NSF and
STScI. STScI is operated by AURA, Inc., under NASA contract NAS 5-26555.

\begin{table*}
 \centering
 \begin{minipage}{200mm}
  \caption{Parameters for Gravitational Potentials}\label{tab:pot}
  \begin{tabular}{@{}lcccccccc@{}}
  \hline
   System & $z$ & $M_\bullet$ & $M_{\rm gal}$ & $R_{\rm e}$
     & $\sigma$ & $c_{\rm vir}$ & $M_{\rm vir}$ & References
\footnote{References for cluster parameters. 
(1) \citet{waj10a}; (2) \citet{vik06a}; 
(3) \citet{pif05a};\\ (4) \citet{ett02b}; (5) \citet{sch07a}; 
(6) \citet{mol99a}; (7) \citet{dav01a};\\ (8) \citet{ric10a}; 
(9) \citet{ett10a}; (10) \citet{mcl99a}; (11) \citet{lok06a};\\
(12) \citet{giz04a}, \citet{sun09a}; (14) \citet{smi02a};
(15) \citet{poi05a}
}
 \\
          &     & $(10^9\: M_\odot)$ & $(10^{11}\: M_\odot)$ 
     & (kpc)
     & $(\rm km\: s^{-1})$ &  & $(10^{14}\: M_\odot)$ &  \\
\hline
A85           & 0.055&
 7.0&
$ 31.0\pm 1.0$&
$ 16.3\pm
 0.03$&
$ 348.4\pm 18.6$&
$ 4.25
^{+ 0.76}_{- 0.96}$&
$12.33
^{+ 1.78}
_{- 1.34}$&
    1
\\
A133          & 0.060&
 3.0&
$ 17.9\pm 0.4$&
$ 14.6\pm
 0.43$&
$ 236.2\pm 11.4$&
$ 6.35
^{+ 0.53}_{- 0.53}$&
$ 5.64
^{+ 0.88}
_{- 0.77}$&
    2
\\
A262          & 0.016&
 0.6&
$  4.9\pm 0.1$&
$ 10.4\pm
 0.58$&
$ 229.8\pm  9.7$&
$ 8.84
^{+ 0.69}_{- 0.69}$&
$ 1.15
^{+ 0.092}
_{- 0.15}$&
    3
\\
Perseus       & 0.018&
 0.34&
$ 19.2\pm 0.1$&
$ 11.3\pm
 0.43$&
$ 258.9\pm 13.4$&
$ 8.08
^{+ 0.35}_{- 0.35}$&
$ 6.81
^{+ 0.63}
_{- 0.72}$&
    4
\\
2A~0335+096   & 0.035&
 3.0&
$ 18.0\pm 1.0$&
$ 15.0\pm
 1.4$&
$ 289.5\pm 35.9$&
$ 7.44
^{+ 0.42}_{- 0.42}$&
$ 2.11
^{+ 0.24}
_{- 0.29}$&
    3
\\
A478          & 0.081&
 5.8&
$ 28.0\pm 1.0$&
$ 15.8\pm
 3.1$&
$ 289.5\pm 35.9$&
$ 5.15
^{+ 0.45}_{- 0.49}$&
$ 16.6
^{+  2.0}
_{-  2.6}$&
    5
\\
MS~0735.6+7421& 0.216&
 5.0&
$ 24.0\pm 1.0$&
$ 15.1\pm
 3.8$&
$ 289.5\pm 35.9$&
$ 8.41
^{+ 8.37}_{- 4.20}$&
$ 25
^{+ 24.9}
_{- 12.5}$&
    6
\\
PKS~0745-191  & 0.103&
 5.5&
$ 27.0\pm 1.0$&
$ 16.1\pm
 3.3$&
$ 289.5\pm 35.9$&
$ 7.75
^{+ 2.15}_{- 1.41}$&
$ 14.9
^{+  6.7}
_{-  3.7}$&
    5
\\
Hydra~A       & 0.055&
 5.8&
$ 28.2\pm 0.7$&
$ 10.5\pm
 0.90$&
$ 361.9\pm 19.1$&
$15.90
^{+ 0.23}_{- 0.23}$&
$ 1.15
^{+ 0.44}
_{- 0.36}$&
    7
\\
Zw~2701       & 0.214&
 6.5&
$ 30.0\pm 1.0$&
$ 13.4\pm
 1.3$&
$ 289.5\pm 35.9$&
$ 3.30
^{+  1.2}_{-  1.2}$&
$10.86
^{+ 2.57}
_{- 5.86}$&
    8
\\
Zw~3146       & 0.291&
 9.0&
$ 13.5\pm 6.9$&
$ 17.4\pm
 7.6$&
$ 289.5\pm 35.9$&
$ 4.19
^{+ 0.18}_{- 0.31}$&
$ 9.29
^{+ 1.04}
_{- 0.55}$&
    9
\\
M84           & 0.0035&
 0.36&
$  4.3\pm 1.3$&
$ 2.45\pm
 0.06$&
$ 282.4\pm  2.6$&
 $\cdots$&
 $\cdots$&
 $\cdots$
\\
M87           & 0.0042&
 6.4&
$ 11.0\pm 3.3$&
$ 3.67\pm
 0.13$&
$ 336.4\pm  4.6$&
$ 3.84
^{+ 0.91}_{- 0.92}$&
$ 5.78
^{+ 0.59}
_{-  1.5}$&
   10
\\
Centaurus     & 0.011&
 2.0&
$ 11.6\pm 0.1$&
$ 9.44\pm
 0.24$&
$ 254.2\pm  7.3$&
$ 7.75
^{+ 0.77}_{- 0.78}$&
$ 4.09
^{+ 0.32}
_{- 0.62}$&
    4
\\
HCG~62        & 0.014&
 0.65&
$ 13.5\pm 6.9$&
$ 6.87\pm
 0.04$&
$ 289.5\pm 35.9$&
 $\cdots$&
 $\cdots$&
 $\cdots$
\\
A1795         & 0.063&
 2.2&
$ 13.4\pm 0.6$&
$ 20.8\pm
 0.23$&
$ 302.0\pm  8.7$&
$ 6.16
^{+ 1.14}_{- 1.14}$&
$ 10.8
^{+  2.7}
_{-  2.4}$&
    5
\\
A1835         & 0.253&
 6.7&
$ 13.5\pm 6.9$&
$ 18.4\pm
 0.35$&
$ 289.5\pm 35.9$&
$ 4.18
^{+ 0.63}_{- 0.41}$&
$ 24.3
^{+  4.4}
_{-  4.9}$&
    5
\\
PKS~1404-267  & 0.022&
 0.7&
$  5.7\pm 0.5$&
$ 6.03\pm
 0.12$&
$ 259.7\pm  6.5$&
$12.25
^{+ 1.09}_{- 6.07}$&
$ 1.77
^{+ 0.43}
_{- 0.31}$&
    1
\\
A2029         & 0.077&
 4.0&
$ 21.9\pm 0.2$&
$ 24.2\pm
 1.6$&
$ 390.8\pm 10.0$&
$ 8.86
^{+ 0.44}_{- 0.50}$&
$ 10.1
^{+ 0.99}
_{- 0.77}$&
    5
\\
A2052         & 0.035&
 2.0&
$ 11.0\pm 3.3$&
$ 15.7\pm
 0.27$&
$ 215.8\pm 11.6$&
$ 6.50
^{+ 0.71}_{- 0.71}$&
$ 2.96
^{+ 0.52}
_{- 0.77}$&
    3
\\
MKW~3S        & 0.045&
 2.0&
$ 11.2\pm 0.3$&
$ 11.6\pm
 2.3$&
$ 289.5\pm 35.9$&
$ 7.83
^{+ 0.55}_{- 0.55}$&
$ 2.90
^{+ 0.27}
_{- 0.38}$&
    3
\\
A2199         & 0.030&
 2.7&
$ 15.7\pm 0.2$&
$ 10.6\pm
 0.20$&
$ 307.1\pm  6.9$&
$10.40
^{+ 14.6}_{-  7.9}$&
$  7.1
^{+  3.4}
_{-  2.4}$&
   11
\\
Hercules~A    & 0.154&
 2.5&
$ 15.0\pm 4.0$&
$ 20.1\pm
 2.0$&
$ 289.5\pm 35.9$&
$ 3.51
^{+ 0.23}_{- 0.23}$&
$ 4.33
^{+ 0.54}
_{- 0.54}$&
12,13
\\
3C~388        & 0.092&
 4.5&
$ 23.0\pm 6.0$&
$ 11.9\pm
 1.2$&
$ 408.3\pm 25.7$&
 $\cdots$&
 $\cdots$&
 $\cdots$
\\
Cygnus~A      & 0.056&
 2.7&
$  9.0\pm 2.0$&
$ 15.6\pm
 0.77$&
$ 289.5\pm 35.9$&
$16.40
^{+ 0.25}_{- 0.25}$&
$ 8.33
^{+ 0.38}
_{- 0.38}$&
13,14
\\
Sersic~159/03 & 0.058&
 2.0&
$ 11.0\pm 2.0$&
$ 20.2\pm
 0.95$&
$ 289.5\pm 35.9$&
$ 8.57
^{+ 0.69}_{- 0.69}$&
$ 1.61
^{+ 0.12}
_{- 0.20}$&
    3
\\
A2597         & 0.085&
 1.5&
$  9.0\pm 1.0$&
$ 11.7\pm
 1.3$&
$ 210.0\pm 57.1$&
$ 7.60
^{+ 0.63}_{- 0.63}$&
$ 3.55
^{+ 0.43}
_{- 0.40}$&
   15
\\
A4059         & 0.048&
 8.7&
$ 38.2\pm 0.4$&
$ 18.7\pm
 0.10$&
$ 272.1\pm 12.9$&
$ 3.57
^{+ 0.68}_{- 0.96}$&
$ 4.45
^{+ 0.63}
_{- 0.62}$&
    1
\\
\hline
\end{tabular}
\end{minipage}
\end{table*}

\begin{table*}
 \centering
 \begin{minipage}{200mm}
  \caption{Observational Data}\label{tab:obs}
  \begin{tabular}{@{}lcccccc@{}}
  \hline
   System & $r_{\rm obs}$ & $n_{\rm e,obs}$ & $T_{\rm obs}$ 
     & $P_{\rm jet}$  & $L_{\rm X}$ & $L_{\rm ICM}$  \\
          &   (kpc)       & $(\rm cm^{-3})$ &    (keV) 
     & $(10^{42}\rm\: erg\: s^{-1})$
     & $(10^{40}\rm\: erg\: s^{-1})$ &  $(10^{42}\rm\: erg\: s^{-1})$  \\
\hline
A85           &  5.8&
$  0.107^{+  0.009}_{-  0.008}$&
$  2.1^{+  0.1}_{-  0.2}$&
$     23^{+     23}_{     -7}$&
$<   15$&
$    335^{+     21}_{    -29 }$
\\
A133          &  8.0&
$  0.048^{+  0.004}_{-  0.005}$&
$  1.8^{+  0.1}_{-  0.1}$&
$    387^{+    162}_{    -13}$&
$< 1.73$&
$    103^{+      3}_{     -3 }$
\\
A262          &  3.4&
$  0.065^{+  0.008}_{-  0.007}$&
$ 0.86^{+ 0.01}_{- 0.01}$&
$  6.1^{+  4.7}_{ -1.6}$&
$< 0.02$&
$ 11.10^{+ 0.32}_{-0.46 }$
\\
Perseus       &  8.6&
$  0.150^{+  0.005}_{-  0.005}$&
$  4.4^{+  0.5}_{-  0.4}$&
$     94^{+     63}_{    -19}$&
$\cdots$&
$    533^{+      7}_{     -8 }$
\\
2A~0335+096   &  5.1&
$  0.056^{+  0.003}_{-  0.002}$&
$  1.4^{+  0.1}_{-  0.1}$&
$     15^{+     14}_{     -4}$&
$< 0.28$&
$    325^{+      5}_{     -4 }$
\\
A478          &  5.3&
$   0.20^{+   0.01}_{-   0.02}$&
$  2.7^{+  0.3}_{-  0.3}$&
$     63^{+     50}_{    -13}$&
$<   23$&
$   1400^{+     23}_{    -51 }$
\\
MS~0735.6+7421& 23.8&
$  0.067^{+  0.002}_{-  0.003}$&
$  3.2^{+  0.2}_{-  0.2}$&
  21900&
$<   17$&
$    438^{+     12}_{    -16 }$
\\
PKS~0745-191  & 11.2&
$   0.14^{+   0.01}_{-   0.01}$&
$  2.6^{+  0.4}_{-  0.4}$&
$   1060^{+    875}_{   -188}$&
$     71^{+      5}_{    -11}$&
$   2070^{+    127}_{   -122 }$
\\
Hydra~A       &  4.7&
$   0.15^{+   0.01}_{-   0.02}$&
$  2.6^{+  0.8}_{-  0.5}$&
$   1250^{+     31}_{    -31}$&
$    105^{+      6}_{     -5}$&
$    269^{+      5}_{     -4 }$
\\
Zw~2701       & 37.6&
$  0.024^{+  0.002}_{-  0.002}$&
$  3.3^{+  0.3}_{-  0.3}$&
$   3750^{+   5560}_{  -2190}$&
$<   40$&
$    430^{+     18}_{    -32 }$
\\
Zw~3146       & 15.0&
$  0.177^{+  0.007}_{-  0.007}$&
$  3.1^{+  0.3}_{-  0.2}$&
$   3620^{+   4250}_{   -938}$&
$<  891$&
$   2330^{+    165}_{   -193 }$
\\
M84           &  0.9&
$  0.105^{+  0.007}_{-  0.007}$&
$ 0.57^{+ 0.01}_{- 0.01}$&
$  0.6^{+  0.9}_{ -0.4}$&
$ 0.41^{+ 0.03}_{-0.08}$&
$  0.06^{+ 0.01}_{-0.01 }$
\\
M87           &  1.0&
$  0.191^{+  0.009}_{-  0.009}$&
$ 0.94^{+ 0.02}_{- 0.02}$&
$  3.8^{+  2.6}_{ -0.6}$&
$ 2.58^{+ 0.16}_{-0.16}$&
$  7.20^{+ 0.20}_{-0.11 }$
\\
Centaurus     &  1.3&
$   0.23^{+   0.01}_{-   0.01}$&
$ 0.77^{+ 0.01}_{- 0.01}$&
$  4.6^{+  3.6}_{ -1.1}$&
$< 0.03$&
$ 23.80^{+ 0.36}_{-0.36 }$
\\
HCG~62        &  2.1&
$  0.057^{+  0.007}_{-  0.005}$&
$ 0.67^{+ 0.01}_{- 0.01}$&
$  2.4^{+  3.8}_{ -1.4}$&
$< 0.01$&
$  1.80^{+ 0.17}_{-0.25 }$
\\
A1795         &  9.5&
$  0.067^{+  0.005}_{-  0.005}$&
$  2.7^{+  0.6}_{-  0.4}$&
$    100^{+    144}_{    -31}$&
$<   15$&
$    615^{+      9}_{    -18 }$
\\
A1835         & 27.2&
$  0.110^{+  0.003}_{-  0.003}$&
$  4.0^{+  0.3}_{-  0.3}$&
$   1120^{+   1190}_{   -375}$&
$<  235$&
$   3160^{+     59}_{    -89 }$
\\
PKS~1404-267  &  8.5&
$  0.046^{+  0.002}_{-  0.002}$&
$  1.3^{+  0.1}_{-  0.1}$&
$     13^{+     16}_{     -6}$&
$     22^{+      1}_{     -1}$&
$     24^{+      1}_{     -1 }$
\\
A2029         &  2.2&
$   0.37^{+   0.04}_{-   0.03}$&
$  2.9^{+  0.3}_{-  0.2}$&
$     54^{+     31}_{     -3}$&
$<   21$&
$   1160^{+      9}_{    -11 }$
\\
A2052         &  5.5&
$  0.017^{+  0.002}_{-  0.002}$&
$ 0.71^{+ 0.04}_{- 0.08}$&
$     94^{+    125}_{     -4}$&
$ 8.82^{+ 0.57}_{-0.85}$&
$     94^{+      1}_{     -1 }$
\\
MKW~3S        &  7.8&
$  0.028^{+  0.006}_{-  0.009}$&
$  2.8^{+  0.8}_{-  0.5}$&
$    256^{+    262}_{    -28}$&
$< 0.95$&
$     99^{+      3}_{     -4 }$
\\
A2199         &  4.4&
$  0.099^{+  0.005}_{-  0.005}$&
$  2.2^{+  0.2}_{-  0.1}$&
$    169^{+    156}_{    -38}$&
$ 1.87^{+ 1.25}_{-1.25}$&
$    142^{+      1}_{     -3 }$
\\
Hercules~A    & 67.0&
$ 0.0111^{+ 0.0006}_{- 0.0005}$&
$  2.0^{+  0.2}_{-  0.2}$&
  10000&
$<   34$&
$    210^{+      6}_{    -52 }$
\\
3C~388        & 55.6&
$ 0.0069^{+ 0.0004}_{- 0.0004}$&
$  3.0^{+  0.2}_{-  0.2}$&
$    125^{+    175}_{    -50}$&
$     43^{+     21}_{    -21}$&
$     27^{+      2}_{     -4 }$
\\
Cygnus~A      &  5.3&
$  0.132^{+  0.009}_{-  0.008}$&
$  5.2^{+  0.5}_{-  0.6}$&
   2440&
$  19200^{+    451}_{   -451}$&
$    370^{+     11}_{    -11 }$
\\
Sersic~159/03 & 12.2&
$  0.056^{+  0.004}_{-  0.004}$&
$  1.8^{+  0.2}_{-  0.1}$&
$    488^{+    512}_{   -162}$&
$< 0.57$&
$    211^{+      8}_{     -8 }$
\\
A2597         & 11.0&
$  0.073^{+  0.005}_{-  0.005}$&
$  1.6^{+  0.2}_{-  0.2}$&
$     42^{+     54}_{    -18}$&
$<   23$&
$    440^{+     20}_{    -36 }$
\\
A4059         & 10.6&
$  0.022^{+  0.001}_{-  0.001}$&
$  2.1^{+  0.1}_{-  0.1}$&
$     60^{+     56}_{    -22}$&
$< 0.44$&
$     91^{+      1}_{     -1 }$
\\
\hline
\end{tabular}
\end{minipage}
\end{table*}

\begin{table*}
 \centering
 \begin{minipage}{200mm}
  \caption{Parameters for the Bondi Accretion}\label{tab:res}
  \begin{tabular}{@{}lcccccc@{}}
  \hline
   System & $r_{\rm B}$ & $n_{\rm e,B}$ & $T_{\rm B}$ 
   & $\dot{M}_{\rm B}$  & $P_{\rm B}$ \\
          &   (kpc)       & $(\rm cm^{-3})$ &    (keV) 
     & $(M_\odot\rm\: yr^{-1})$ & $(10^{44}\rm\: erg\: s^{-1})$  \\
\hline
A85           &
$  0.15^{+  0.31}_{ -0.10}$&
$  2.73^{+  0.73}_{ -0.81}$&
$ 1.56^{+ 0.17}_{-0.15}$&
$  0.85^{+  5.60}_{ -0.75}$&
$     48^{+    318}_{    -43}$&
\\
A133          &
$  0.13^{+  0.27}_{ -0.09}$&
$  5.68^{+  1.92}_{ -2.16}$&
$ 0.73^{+ 0.08}_{-0.06}$&
$  1.01^{+  5.42}_{ -0.89}$&
$     58^{+    307}_{    -51}$&
\\
A262          &
$ 0.029^{+ 0.063}_{-0.020}$&
$  0.76^{+  0.20}_{ -0.15}$&
$ 0.67^{+ 0.06}_{-0.05}$&
$ 0.006^{+ 0.052}_{-0.005}$&
$ 0.35^{+ 2.93}_{-0.31}$&
\\
Perseus       &
$ 0.013^{+ 0.028}_{-0.009}$&
$    34^{+    13}_{    -9}$&
$ 0.86^{+ 0.09}_{-0.08}$&
$ 0.062^{+ 0.513}_{-0.055}$&
$  3.5^{+ 29.1}_{ -3.1}$&
\\
2A~0335+096   &
$ 0.092^{+ 0.203}_{-0.062}$&
$  1.27^{+  0.97}_{ -0.52}$&
$ 1.07^{+ 0.27}_{-0.24}$&
$  0.13^{+  1.13}_{ -0.11}$&
$  7.2^{+ 64.0}_{ -6.5}$&
\\
A478          &
$  0.17^{+  0.33}_{ -0.11}$&
$  7.93^{+ 10.16}_{ -4.22}$&
$ 1.12^{+ 0.29}_{-0.21}$&
$  2.78^{+ 17.71}_{ -2.48}$&
$    158^{+   1000}_{   -141}$&
\\
MS~0735.6+7421&
$  0.15^{+  0.32}_{ -0.10}$&
$    74^{+  3436}_{   -57}$&
$ 1.09^{+ 0.28}_{-0.22}$&
$ 20.07^{+1572.5}_{-18.50}$&
$   1140^{+  89100}_{  -1050}$&
\\
PKS~0745-191  &
$  0.17^{+  0.35}_{ -0.11}$&
$    25^{+    70}_{   -15}$&
$ 1.09^{+ 0.28}_{-0.22}$&
$  8.23^{+ 84.95}_{ -7.43}$&
$    466^{+   4810}_{   -421}$&
\\
Hydra~A       &
$  0.11^{+  0.24}_{ -0.08}$&
$    11^{+     6}_{    -5}$&
$ 1.69^{+ 0.19}_{-0.15}$&
$  2.10^{+ 12.07}_{ -1.86}$&
$    119^{+    684}_{   -106}$&
\\
Zw~2701       &
$  0.19^{+  0.40}_{ -0.13}$&
$    36^{+    57}_{   -24}$&
$ 1.10^{+ 0.28}_{-0.22}$&
$ 16.53^{+105.51}_{-15.13}$&
$    937^{+   5980}_{   -858}$&
\\
Zw~3146       &
$  0.27^{+  0.51}_{ -0.18}$&
$  5.72^{+ 19.59}_{ -3.37}$&
$ 1.11^{+ 0.31}_{-0.20}$&
$  4.89^{+ 57.30}_{ -4.39}$&
$    277^{+   3250}_{   -249}$&
\\
M84           &
$ 0.012^{+ 0.026}_{-0.008}$&
$  9.86^{+ 25.29}_{ -7.22}$&
$ 1.01^{+ 0.02}_{-0.02}$&
$ 0.016^{+ 0.171}_{-0.015}$&
$ 0.88^{+ 9.67}_{-0.83}$&
\\
M87           &
$  0.15^{+  0.42}_{ -0.11}$&
$  8.16^{+ 14.66}_{ -7.32}$&
$ 1.36^{+ 0.06}_{-0.22}$&
$  2.60^{+  3.58}_{ -2.34}$&
$    148^{+    203}_{   -133}$&
\\
Centaurus     &
$ 0.080^{+ 0.175}_{-0.055}$&
$  2.19^{+  0.38}_{ -0.67}$&
$ 0.82^{+ 0.04}_{-0.04}$&
$  0.15^{+  0.90}_{ -0.13}$&
$  8.3^{+ 50.9}_{ -7.3}$&
\\
HCG~62        &
$ 0.020^{+ 0.046}_{-0.014}$&
$  4.16^{+ 36.99}_{ -3.63}$&
$ 1.06^{+ 0.27}_{-0.25}$&
$ 0.020^{+ 0.497}_{-0.019}$&
$  1.1^{+ 28.2}_{ -1.1}$&
\\
A1795         &
$ 0.062^{+ 0.130}_{-0.042}$&
$  1.17^{+  0.23}_{ -0.20}$&
$ 1.17^{+ 0.07}_{-0.06}$&
$ 0.055^{+ 0.442}_{-0.049}$&
$  3.1^{+ 25.0}_{ -2.8}$&
\\
A1835         &
$  0.20^{+  0.41}_{ -0.13}$&
$  7.00^{+  8.50}_{ -3.90}$&
$ 1.10^{+ 0.28}_{-0.22}$&
$  3.36^{+ 26.44}_{ -3.06}$&
$    191^{+   1500}_{   -173}$&
\\
PKS~1404-267  &
$ 0.027^{+ 0.057}_{-0.018}$&
$  9.02^{+  3.47}_{ -4.06}$&
$ 0.86^{+ 0.04}_{-0.04}$&
$ 0.068^{+ 0.489}_{-0.063}$&
$  3.9^{+ 27.7}_{ -3.5}$&
\\
A2029         &
$ 0.067^{+ 0.137}_{-0.045}$&
$  1.72^{+  0.28}_{ -0.27}$&
$ 1.97^{+ 0.12}_{-0.09}$&
$  0.12^{+  0.92}_{ -0.11}$&
$  7.0^{+ 52.2}_{ -6.2}$&
\\
A2052         &
$  0.11^{+  0.24}_{ -0.08}$&
$  0.53^{+  0.47}_{ -0.27}$&
$ 0.60^{+ 0.06}_{-0.06}$&
$ 0.057^{+ 0.437}_{-0.052}$&
$  3.2^{+ 24.8}_{ -2.9}$&
\\
MKW~3S        &
$ 0.061^{+ 0.130}_{-0.041}$&
$  1.02^{+  1.11}_{ -0.52}$&
$ 1.08^{+ 0.28}_{-0.23}$&
$ 0.045^{+ 0.403}_{-0.041}$&
$  2.5^{+ 22.8}_{ -2.3}$&
\\
A2199         &
$ 0.073^{+ 0.150}_{-0.049}$&
$  3.81^{+  5.45}_{ -1.17}$&
$ 1.22^{+ 0.07}_{-0.05}$&
$  0.26^{+  2.31}_{ -0.22}$&
$     15^{+    131}_{    -13}$&
\\
Hercules~A    &
$ 0.077^{+ 0.173}_{-0.052}$&
$  1.28^{+  2.01}_{ -0.70}$&
$ 1.07^{+ 0.27}_{-0.25}$&
$ 0.089^{+ 1.032}_{-0.082}$&
$  5.1^{+ 58.5}_{ -4.7}$&
\\
3C~388        &
$ 0.069^{+ 0.151}_{-0.047}$&
$  0.65^{+  1.23}_{ -0.41}$&
$ 2.13^{+ 0.27}_{-0.26}$&
$ 0.053^{+ 0.542}_{-0.049}$&
$  3.0^{+ 30.7}_{ -2.8}$&
\\
Cygnus~A      &
$ 0.078^{+ 0.151}_{-0.052}$&
$  3.39^{+  1.23}_{ -1.07}$&
$ 1.13^{+ 0.30}_{-0.21}$&
$  0.25^{+  1.58}_{ -0.23}$&
$     14^{+     90}_{    -13}$&
\\
Sersic~159/03 &
$ 0.061^{+ 0.138}_{-0.042}$&
$  1.02^{+  0.67}_{ -0.37}$&
$ 1.07^{+ 0.27}_{-0.24}$&
$ 0.046^{+ 0.447}_{-0.041}$&
$  2.6^{+ 25.3}_{ -2.4}$&
\\
A2597         &
$ 0.086^{+ 0.210}_{-0.060}$&
$    13^{+    47}_{    -9}$&
$ 0.57^{+ 0.33}_{-0.24}$&
$  0.84^{+ 17.18}_{ -0.80}$&
$     48^{+    974}_{    -45}$&
\\
A4059         &
$  0.29^{+  0.58}_{ -0.20}$&
$  4.28^{+  1.74}_{ -2.13}$&
$ 0.98^{+ 0.11}_{-0.08}$&
$  4.14^{+ 15.65}_{ -3.57}$&
$    235^{+    887}_{   -202}$&
\\
\hline
\end{tabular}
\end{minipage}
\end{table*}

\end{document}